\documentclass[aps,pra,twocolumn,superscriptaddress,floatfix,citeautoscript,preprintnumbers]{revtex4-1}
\usepackage[dvipsnames]{xcolor}
\usepackage{amsmath,amssymb}
\usepackage{commath}
\usepackage{todonotes}
\usepackage{siunitx}
\usepackage[colorlinks = true,
            linkcolor = blue,
            urlcolor  = blue,
            citecolor = blue,
            anchorcolor = blue]{hyperref}
            
\usepackage{cleveref}
\usepackage{graphicx}
\begin{document}
\crefname{equation}{Eq.}{Eqs.}
\crefname{figure}{Fig.}{Fig.}
\crefname{appendix}{Appendix}{Appendix}

\newcommand{\dan}[1]{\textcolor{BrickRed}{#1}}

\title{Generation of photonic tensor network states with Circuit QED}

\author{Zhi-Yuan Wei}
\email{zhiyuan.wei@mpq.mpg.de}
\author{J. Ignacio Cirac}
\author{Daniel Malz}
\email{daniel.malz@mpq.mpg.de}

\affiliation{Max-Planck-Institut f{\"u}r Quantenoptik, Hans-Kopfermann-Stra{\ss}e 1, D-85748 Garching, Germany}
\affiliation{Munich Center for Quantum Science and Technology (MCQST), Schellingstr. 4, D-80799 M{\"u}nchen, Germany}

\date{\today}

\begin{abstract}
We propose a circuit QED platform and protocol to generate microwave photonic tensor network states deterministically. We first show that using a microwave cavity as ancilla and a transmon qubit as emitter is a good platform to produce photonic matrix product states. The ancilla cavity combines a large controllable Hilbert space with a long coherence time, which we predict translates into a high number of entangled photons and states with a high bond dimension.
Going beyond this paradigm, we then consider a natural generalization of this platform, in which several cavity--qubit pairs are coupled to form a chain.
The photonic states thus produced feature a two-dimensional entanglement structure and can be interpreted as \textit{radial plaquette} projected entangled pair states [Z.Y. Wei, D. Malz, and J. I. Cirac., Phys. Rev. Lett. 128, 010607 (2022)], which include many paradigmatic states, such as the broad class of isometric tensor network states, graph states and string-net states.

\end{abstract}

\maketitle

\section{Introduction}
Producing large-scale entangled photonic states is central to many quantum technologies, including computing~\cite{OBrien2009}, cryptography~\cite{gisin2007quantum}, networks~\cite{kimble2008quantum}, or sensing~\cite{Degen2017}. The standard method for producing multi-photon entanglement utilizes parametric down-conversion (PDC)~\cite{burnham1970observation}, which has been used to produce 12-photon entanglement~\cite{Zhong2018}.
However, PDC possesses certain limitations, most notably the exponential decrease of success probability with photon number.
One promising way to overcome that is to deterministically and sequentially generate a string of entangled photons using a single quantum emitter~\cite{gheri1998entanglement,saavedra2000controlled,Schon2005,Schon2007,Lindner2009,Tiurev2020a,zyryd}. 
The class of states that can thus be generated coincides with the set of matrix product states (MPS)~\cite{Schon2005}, a type of tensor network states (TNS) that widely appears in one-dimensional quantum many-body systems~\cite{Perez-Garcia2006,schollwock2011density,orus2014practical,Cirac2020}. Some sequential photon generation protocols have been experimentally realized in quantum dots~\cite{schwartz2016} and circuit QED~\cite{Eichler2015,Besse2020}.
Using coupled emitters~\cite{Economou2010,Gimeno-Segovia2019,Russo2019,Bekenstein,Bartolucci2021} or allowing the emitted photons to travel back and interact with the photon source again~\cite{Pichler2017, Dhand2018,Xu2018,Wan2020,Zhan2020,Shi2021,Bombin2021}, it is possible to produce certain projected entangled-pair states (PEPS)~\cite{verstraete2004renormalization}, which are higher-dimensional generalization of MPS.

Preparing most PEPS is known to be difficult, as they generally require a preparation time that increases exponentially with the system size ~\cite{Verstraete2006,Schuch2007}. 
In contrast, a broad subset of PEPS that can be prepared efficiently are the \textit{radial plaquette} PEPS (rp-PEPS)~\cite{zypp}. These are obtained 
 through the sequential application of geometrically local unitaries in the form of plaquettes of side length $L_p$.
rp-PEPS contain isometric tensor network states (isoTNS) as a subclass~\cite{Zaletel2020}, which are PEPS~\cite{Perez-Garcia2006,schollwock2011density,orus2014practical,Cirac2020} subject to an isometry condition. This immediately implies that rp-PEPS include important states such as the graph states with local connectivities~\cite{PhysRevA.69.062311,Russo2019}, toric codes \cite{Verstraete2006,Schuch2010}, all string-net states~\cite{levin2005,gu2009tensor,Buerschaper2009,Soejima2020}~\footnote{Note that the string-net states are originally defined on the hexagonal lattice~\cite{levin2005}. Since one can embed a hexagonal lattice into a square lattice, one can create string-net states using the schemes we show in this paper, which are based on square lattices.} and hypergraph states with local connectivities~\cite{Takeuchi2019}.
The experimental preparation of such states in two dimensions is pursued intensely~\cite{Asavanant2019,satzinger2021,semeghini2021probing}.

To date, existing platforms and proposals have almost exclusively explored photonic MPS of bond dimension $D=2$ \cite{gheri1998entanglement,saavedra2000controlled,Lindner2009,Tiurev2020a,schwartz2016,Besse2020}, with one theoretical protocol forming an exception, which is capable of deterministically producing MPS with higher bond dimensions using an ordered array of Rydberg atoms~\cite{zyryd}. These platforms, however, do not easily extend to produce higher-dimensional PEPS.
The existing proposals that produce higher-dimensional PEPS are also mostly limited to $D=2$, and particularly focus on cluster state generation. One notable exception is the protocol in Ref.~\cite{Dhand2018}, which produces two-dimensional PEPS with $D > 2$ utilizing the PDC process in an optical loop, but with a probablistic protocol. Thus, despite significant efforts, there are still important theoretical challenges on deterministically producing high-fidelity and high-bond-dimension photonic tensor network states in one and particularly in higher dimensions. 
Sequential generation of photonic tensor network states with high bond dimension would allow creating states useful for quantum metrology ~\cite{Jarzyna2013,Chabuda2020}, ancilla-photon superposed states~\cite{zyryd} useful for quantum networks~\cite{miguel2021genuine}, and ground states of a large variety of many-body systems useful for quantum simulation~\cite{orus2014practical, Huang2015a,Dalzell2019,Huang2019a,Schuch2017}.

In this work, we propose a circuit QED platform capable of deterministically generating (microwave) photonic rp-PEPS with the so-called \textit{source point}~\cite{zypp} in one corner of the lattice.
We first consider a cavity dispersively coupled to a transmon qubit and show that this allows one to generate MPS of moderately high bond dimension and many entangled photons, which is outstanding among currently available platforms. Our simulations indicate that this platform has the potential to deterministically generate a one-dimensional cluster state of with a large number of photons using current technologies, which would improve the experimental results in Ref.~\cite{Besse2020} severalfold. 
We then show that by using an array of $m$ such MPS sources, one can efficiently generate rp-PEPS.
The circuit depth in terms of plaquette unitaries to prepare such a state on a $n\times m$ lattice of photons asymptotically scales as~\cite{zypp}
\begin{equation} \label{circ_dep}
{\cal T} \approx L_p \cdot n+m.
\end{equation}
\begin{figure*}[t]
	\centering
	\includegraphics[width=0.9\textwidth]{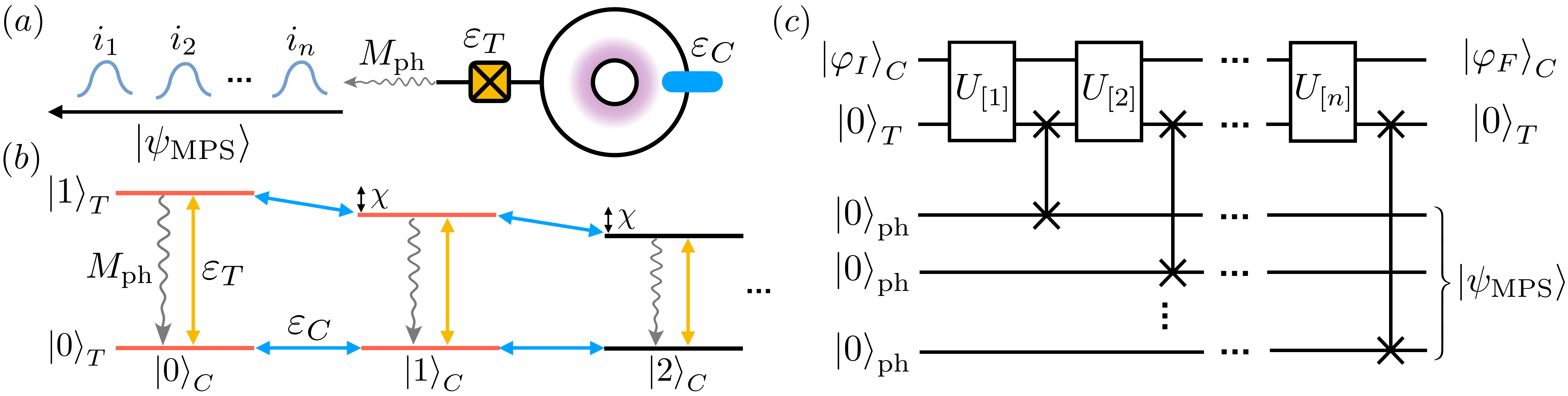}
        \caption{Generating photonic matrix product states (MPS) with circuit QED. (a) We consider a cavity dispersively coupled to a transmon, which can be controlled by driving both the cavity ($\varepsilon_C$) and the transmon ($\varepsilon_T$). The transmon excitation can be converted into a traveling microwave photon. By sequentially applying unitary operations followed by photon emissions $M_{\rm ph}$, one can produce a chain of entangled photons. (b) The level diagram of the system. The red (gray) levels are used in the protocol for $D = 2$ MPS generation. The yellow (light gray) arrows denote driving of the transmon, and the blue (dark gray) arrows denote driving of the cavity. (c) Quantum circuit of the MPS generation protocol. The SWAP gates correspond to photon emissions ($M_{\rm ph}$).}
        \label{fig1}
\end{figure*}
Since rp-PEPS is a large class, our platform allows one to create photonic states useful for applications in quantum computing~\cite{PhysRevA.69.062311,Takeuchi2019}, metrology~\cite{Jarzyna2013,Koczor2020,Chabuda2020}, communication and networking~\cite{Azuma2015}, as well as states that exhibit topological order~\cite{Soejima2020,Verstraete2006,Buerschaper2009,Schuch2010}. Moreover, the sequential nature of the protocol leads to temporally seperated microwave photons, which would allow one to efficiently \textit{distribute} them to multiple receivers, thus directly building a multi-party entangled state. Realization of state transfer and distribution between superconducting quantum processors are intensely pursued currently~\cite{Axline2018,Kurpiers2018,Magnard2020}.

The rest of the article is structured as follows. In \cref{sec_mps}, we present our setup to generate arbitrary MPS using a microwave cavity coupled to a transmon qubit, discuss the imperfections that arise during the MPS generation, and estimate the performance of the device. In \cref{hd_iso}, we present the setup to generate photonic rp-PEPS and provide the circuits for generating the two-dimensional cluster state, the toric code, and isometric tensor network states. Finally, we analyze the scaling of the state preparation fidelity. We summarize our work in \cref{conclu}.

\section{Generating MPS with cQED}
\label{sec_mps}

In this section, first we introduce the setup in \cref{cq_mps_set} and the MPS generation protocol in \cref{cq_mps_prot}. Then we analyze the imperfections during the protocol in \cref{model_err}, and show in \cref{mps_perf} that this protocol can be implemented using current technology.

\subsection{cQED sequential photon source}
\label{cq_mps_set}
We consider the setup sketched in \cref{fig1}(a), where a cavity (with the Hilbert space ${\cal H}_C$) is dispersively coupled to a transmon qubit (with the Hilbert space ${\cal H}_T$), with a Hilbert space ${{\cal H}_{{\rm{src}}}} = {{\cal H}_T} \otimes {{\cal H}_C}$~\cite{Reagor2016}. The transmon ground (excited) state is denoted by ${\left| 0 \right\rangle _T}$ (${\left| 1 \right\rangle _T}$). Defining the transition operator ${\sigma _{\alpha \beta }} = {\left| \alpha  \right\rangle _T}\left\langle \beta  \right|$ for the transmon, the system Hamiltonian $H_{\rm src} (t) = H_0 + H_{\rm drive} (t)$ contains a static part
\begin{equation} \label{mps_ham_dri}
H_{0}=\omega_{T} \sigma_{11}+\omega_{C} a^{\dagger} a-\chi \sigma_{11} a^{\dagger} a,	
\end{equation}
and time-dependent driving of transmon and cavity
\begin{equation} \label{mps_ham_ctl}
	H_{\text {drive }}(t)=\varepsilon_{C}(t) a+\varepsilon_{T}(t) \sigma_{01}+\text { H.c. }
\end{equation}
Here $\omega_T$ $(\omega_C)$ is the frequency of the transmon qubit (cavity), and $a$ is the lowering operator of the cavity mode. The dispersive interaction strength $\chi$ sets a timescale for cavity-transmon gates.
The driving amplitude of the qubit (cavity) is $\varepsilon_T$ $(\varepsilon_C)$. The level structure of this system is shown in \cref{fig1}(b)~\footnote{Here we neglected the higher order Hamiltonian terms such as the Kerr non-linearity~\cite{Heeres2017}. In principle, one can also include these terms in the pulse optimization.}. 
This Hamiltonian gives universal control of the cavity-transmon system~\cite{Strauch2012a,Krastanov2015a}.
We assume that one can engineer the following on-demand photon emission process $M_{\rm ph}$ from the transmon excitation
\begin{equation} \label{ap_map}
	M_{\rm ph}: \quad|i\rangle_{T} \rightarrow|0\rangle_{T}|i\rangle_{\mathrm{ph}}, \quad i=0,1.
\end{equation}
For example, one can realize $M_{\rm ph}$ by coupling the qubit to an emitter via a tunable coupler~\cite{Besse2020}.

\subsection{MPS generation protocol}
\label{cq_mps_prot}
The setup in \cref{cq_mps_set} can sequentially generate photonic MPS using the generic protocol proposed in Ref.~\cite{Schon2005}, schematically shown in \cref{fig1}(c). We identify the first $D$ Fock states of the cavity mode as our basis for the $D$-level ancilla, with a Hilbert space $\cal{H}_D$. The MPS generation protocol starts from an ancilla initial state $\left|\varphi_{I}\right\rangle_{C} \in \mathcal{H}_{D}$ with the transmon in its ground state. In each photon generation round, the ancilla first interacts with the transmon, described by a unitary operation $U_{[i]}$. Then, the transmon emits its excitation and returns to its ground state (denoted by a SWAP gate in \cref{fig1}c), generating a photonic qubit defined by the presence or absence of a photon at that time.

Compared to the setup in Ref.~\cite{Schon2007} which proposes to employ a $D$-level atom as ancilla and an optical cavity as emitter, our setup is better suited to current circuit QED experiments. In particular, our protocol exploits the long coherence times that can be achieved in microwave cavities and uses transmon qubits only to control and emit, but not to store excitations.


Throughout the protocol, unitaries are applied only after emission of the photon from the emitter, such that they always act on ancilla-transmon states of form $|\varphi\rangle_{C}|0\rangle_{T}$. As a result we can specify the action of the unitaries in terms of isometries $V$ that act only on the cavity Hilbert space
\begin{equation} \label{u_act}
U_{[i]}\left(|\varphi\rangle_{C}|0\rangle_{T}\right)=\sum_{j=0}^{1} (V_{[i]}^{j}|\varphi\rangle_{C}) |j\rangle_{T},
\end{equation}
Since $U_{[i]}$ is unitary, the matrices $\{V_{[i]}^{j}\}$ satisfy the isometry condition $\sum_{j=0}^{1} V_{[i]}^{j \dagger} V_{[i]}^{j}=I_{D}$. The quantum state after $n$ rounds of  photon generation is $|\Psi\rangle=M_{\rm ph} U_{[n]} \ldots M_{\rm ph} U_{[1]}\left|\varphi_{I}\right\rangle_{C}$. By disentangling the ancilla and the photonic state in the last step, we obtain the final state $|\Psi\rangle=\left|\varphi_{F}\right\rangle_{C} \otimes\left|\psi_{\mathrm{MPS}}\right\rangle$, which includes the following photonic MPS~\cite{Schon2005}
\begin{equation} \label{mps}
\left|\psi_{\mathrm{MPS}}\right\rangle \propto \sum_{i_{1} \ldots i_{n}=0}^{1} {_C}\langle\varphi_{F}|V_{[n]}^{i_{n}} \ldots V_{[1]}^{i_{1}}| \varphi_{I}\rangle_{C}\left|i_{n} \ldots i_{1}\right\rangle.
\end{equation}
We use the quantum optimal control (QOC) \cite{zyryd,Machnes2018} to find the pulse sequences that implement the desired unitary operations for our protocol (see details in \cref{goat_imple}). As a demonstration, we show how to generate a linear cluster state~\cite{Briegel2001}, which can be written as an MPS of bond dimension $D = 2$, where
\begin{equation} \label{cls_form}
	\begin{array}{l}
V_{[i]}^{0}=\frac{1}{\sqrt{2}}\left(\begin{array}{ll}
1 & 0 \\
1 & 0
\end{array}\right), \quad V_{[i]}^{1}=\frac{1}{\sqrt{2}}\left(\begin{array}{cc}
0 & 1 \\
0 & -1
\end{array}\right), \\
\left|\varphi_{I}\right\rangle_{C}=\frac{1}{\sqrt{2}}(|0\rangle+|1\rangle), \quad\left|\varphi_{F}\right\rangle_{C}=|0\rangle.
\end{array}
\end{equation}

\begin{figure}[htb]
	\centering
	\includegraphics[width=0.48\textwidth]{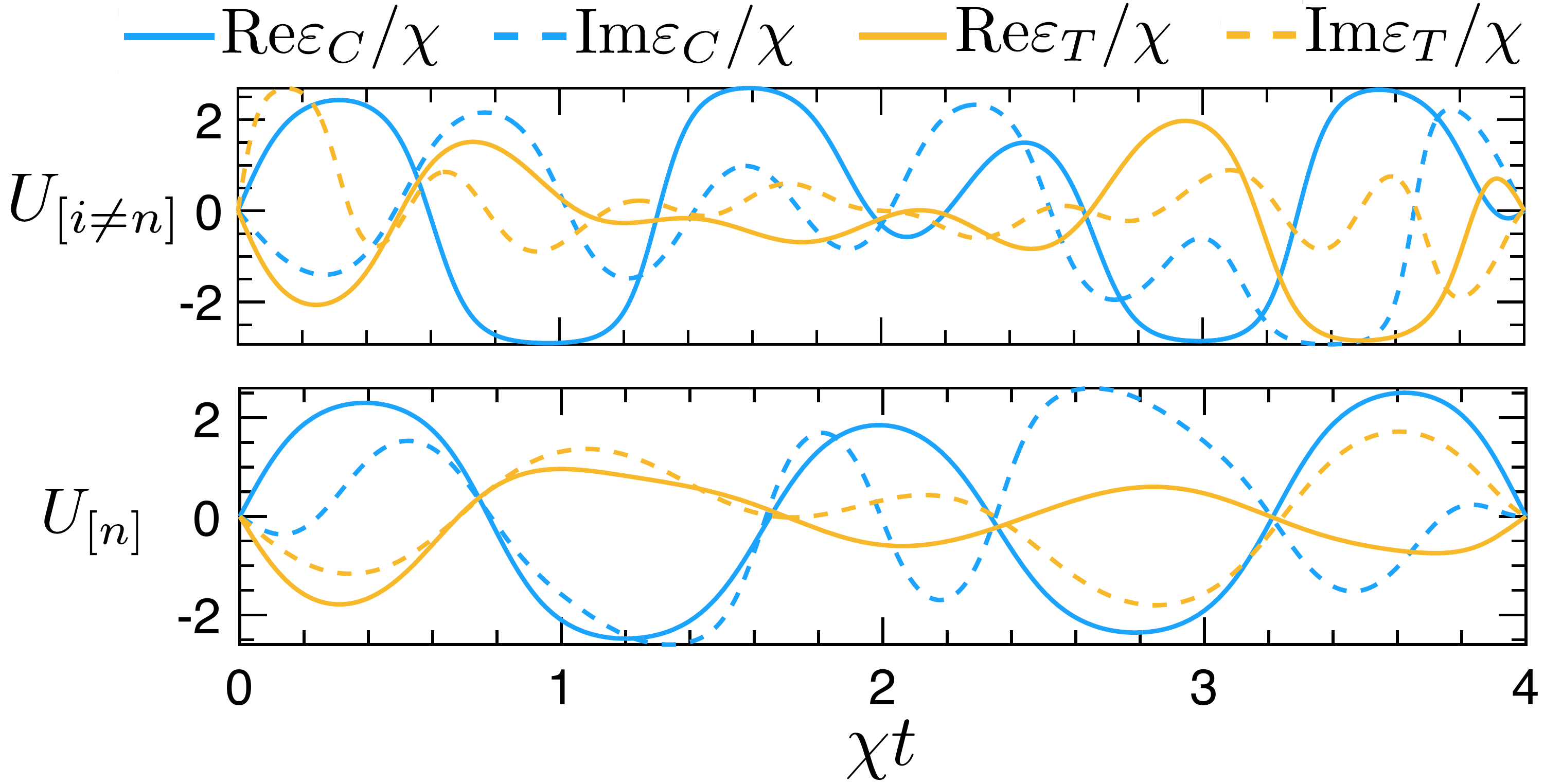}
        \caption{Optimized cavity and transmon drive [\cref{mps_ham_ctl}] to implement the unitaries $U_{[i \neq n]}$ and $U_{[n]}$ for the cluster state [\cref{cls_form}] generation.}
        \label{qoc_cq}
\end{figure}

In each round except the last, we apply the same unitary $U_{[i \neq n]}$ followed by photon emission represented by $M_{\rm ph}$, which adds one site to the state. In the last step, we apply the unitary $U_{[n]}$ followed by $M_{\rm ph}$, which emits the last photon and disentangles the source from the photons. The pulse sequence of the driving [\cref{mps_ham_ctl}] for implementing the two unitaries is shown in \cref{qoc_cq}, and we provide more details in \cref{goat_imple}.

\subsection{Analysis of experimental imperfections}
\label{model_err}
Ideally, the above protocol would generate the desired pure photonic state $\left| {{\psi _{\rm MPS}}} \right\rangle $. However, various errors may occur in this system during the unitary operation and the photon emission process. Thus the protocol produces a $n$-photon density matrix $\rho_{\rm ph}$, with a non-unit fidelity ${{\cal F}_{{\rm{MPS}}}}= {}_{\rm ph}\langle {{\psi _{{\rm{MPS}}}}} |{\rho _{{\rm{ph}}}}{| {{\psi _{{\rm{MPS}}}}} \rangle _{{\rm{ph}}}}$. Due to the sequential nature of the protocol, ${\mathcal{F}_{\rm MPS}}$ is an exponentially decaying function of the emitted photon number $n$, that is 
\begin{equation} \label{err_scale}
{{\cal F}_{{\rm{MPS}}}} = {e^{- \xi  \cdot n}},
\end{equation}
where $\xi$ is the error per photon emission. An example of this behavior is shown in \cref{pe_plot}(a).

Decoherence processes in the cavity-transmon system include transmon decay at a rate ${\Gamma _T}$, cavity mode decay at a rate ${\Gamma _C}$, and transmon dephasing at a rate ${\Gamma _\phi }$ \cite{Heeres2017} \footnote{Note that the thermal excitations
(photon gain) in the cavity and the transmon are neglected here}. These processes happen both during the unitary operations and the photon emission process. The finite anharmonicity $\alpha$ of the transmon further allows leakage into the second excited state $| 2 \rangle_T $ in every unitary operation.

To model the imperfections due to finite anharmonicity $\alpha$ during unitary operations, we model the transmon as a truncated anharmonic oscillator with basis $\{{{{| 0 \rangle }_T},{{| 1 \rangle }_T},{{| 2 \rangle }_T}} \}$. After further including the decoherence effects, the system density matrix ${\rho _{{\rm{src}}}} \in {\cal H}_{\rm src}$ evolves under the master equation
\begin{equation} \label{ext_me}
\begin{aligned}
{{\dot \rho }_{{\rm{src}}}}( t ) &=  -i[ {{H'_{{\rm{src}}}}( t ) ,{\rho _{{\rm{src}}}}( t )} ]\\
& + \sum\limits_{n=T,C,\phi} {( {{J_n}{\rho _{{\rm{src}}}}( t )J_n^\dag  - \frac{1}{2}\{ {{\rho _{{\rm{src}}}}( t ),J_n^\dag {J_n}} \}} )} ,
\end{aligned}
\end{equation}
with $H'_{{\rm{src}}}(t)$ being the system Hamiltonian including transmon double excitations with anharmonicity $\alpha$. Defining the spin operators ${\sigma _{ij}} = {| i \rangle _T}\langle j |$, the jump operators are ${J_T} = \sqrt {{\Gamma _T}} ( {{\sigma _{01}} + \sqrt 2 {\sigma _{12}}} ),{J_C} = \sqrt {{\Gamma _C}} a$, ${J_\phi } = \sqrt {{\Gamma _\phi }} ( {{\sigma _{11}} + 2{\sigma _{22}}} )$.

Since current experiments generally have $| \alpha  |  \gg | \chi  |  \gg {\Gamma _C},{\Gamma _T},{\Gamma _\phi }$ \cite{Reagor2016, Heeres2017,Chou2018}, for a gate time $T\sim1/| \chi  | $ we can estimate the scale of the errors on ${{\cal F}_{{\rm{MPS}}}}$ during each unitary operation perturbatively. The total error $\xi$ is a sum of several parts: (i) transmon decay ${\xi _{{\Gamma _T}}}$, (ii) transmon dephasing ${\xi _{{\Gamma _{\phi}}}}$, (iii) cavity decay ${\xi _{{\Gamma _C}}}$, and (iv) transmon non-linearity ${\xi _{{\alpha }}}$.
These contributions scale as
\begin{equation} \label{quali_unit}
\begin{array}{*{20}{l}}
{{\xi _{{\Gamma _T}}}\sim{\Gamma _T}/| \chi  | ,\qquad {\xi _{{\Gamma _\phi }}}\sim{\Gamma _\phi }/| \chi  | ,}\\
{{\xi _{{\Gamma _C}}}\sim{\Gamma _C}/| \chi  | ,\qquad{\xi _\alpha }\sim{|\chi| ^2}/{| \alpha  | ^2}.}
\end{array}
\end{equation}

Imperfections also affect the photon emission process. First, there is a finite photon retrieval efficiency $p_{\rm em}$ of the emitter~\footnote{The photon retrieval efficiency $p_{\rm em}$ depends on the explicit realization of the on-demand photon emission process, which we do not specify in this work. For example, in Ref.~\cite{Besse2020} the system is modeled with $p_{\rm em}=1$.} associated to $M_{\rm ph}$. Second, system decoherence happens during the photon emission. We assume a finite photon emission rate $\Gamma _{\rm em}$, and thus a finite duration of photon emission ${T_{\rm em}}$ (which we can tune).
In the regime of ${\Gamma _T},{\Gamma _\phi },{\Gamma _C} \ll {\Gamma _{\rm em}}$, we can estimate scaling of error on ${\cal F}_{{{\rm MPS}}}$ due to system decoherences during each emission process as
\begin{equation} \label{quali_emit}
\begin{array}{l}
\xi _{{\Gamma _C}}^{{\rm{em}}}\sim{\Gamma _C} T_{\rm em},\qquad \xi _{{\Gamma _T}}^{{\rm{em}}}\sim{\Gamma _T}/{\Gamma _{{\rm{em}}}},\\
\xi _{{\Gamma _\phi }}^{{\rm{em}}}\sim{\Gamma _\phi }/{\Gamma _{{\rm{em}}}},\qquad{\xi _{{p_{\rm em}}}}\sim -\log {p_{\rm em}}.
\end{array}
\end{equation}
Notably, here $\xi _{{\Gamma _T}}^{\rm em}$ and $\xi _{{\Gamma _{\phi}}}^{\rm em}$, which are due to transmon decoherence, do not depend on the emission time $T_{\rm em}$.
The expression for $\xi _{{p_{\rm em}}}$ comes from the fact that ${e^{ - {\xi _{{p_{{\rm{em}}}}}}{n_{{\rm{ph}}}}}}\sim p_{{\rm{em}}}^{{n_{{\rm{ph}}}}}$. The finite photon emission time also results in a residual population of the transmon first excited state ${p_{1T}}( T_{\rm em} ) = {e^{-\Gamma_{\rm em} {T_{\rm em}}}} \cdot {p_{1T}}( 0 )$, which reduces the photon retrieval efficiency $p_{\rm em}$. The whole photon emission process including all imperfections can be described by a process map ${W_{\rm ph}}:{\cal H}_{\rm src} \to {\cal H}_{\rm src} \otimes {{\cal H}_{\rm ph}}$ that maps ${\rho _{\rm src}}$ to a system-photon joint density matrix (see the construction of $W_{\rm ph}$ in \cref{apd_errors}). Finally, note that we did not include the imperfections during photon transmission, which is not a part of our model.

Using the solution of \cref{ext_me} and the process map ${W_{\rm ph}}$, we can use a matrix product density operator (MPDO) approach~\cite{zyryd} to obtain the photonic state fidelity ${\mathcal{F}_{\rm MPS}}$ and extract the overall error rate $\xi$ (details in \cref{mpdo_fid}). As a demonstration, we analyze the process of generating the one-dimensional cluster state using the pulses in \cref{qoc_cq}. From the scaling data of $\xi$ as a function of various imperfections (details in \cref{beta_scale}), in the regime of small error ($\xi \ll 1$), we have
\begin{equation} \label{fid_scale}
\xi  \approx {\xi _{{\rm{unit}}}} + \xi _{{\rm{em}}}^{{\rm{src}}} + \xi _{{\rm{em}}}^{{\rm{ph}}},
\end{equation}
which includes imperfections during unitary operation
\begin{equation} \label{fid_unit}
{\xi _{{\rm{unit}}}} = {\beta _0} + \frac{{{\beta _C}{\Gamma _C} + {\beta _T}{\Gamma _T} + {\beta _\phi }{\Gamma _\phi }}}{{|\chi |}} + \frac{{{\beta _\alpha } \cdot |\chi {|^2}}}{{|\alpha {|^2}}},
\end{equation}
cavity-transmon decoherence during photon emission
\begin{equation} \label{fid_src_em}
	\xi _{{\rm{em}}}^{{\rm{src}}} =  {{\beta _{\phi ,{\rm{em}}}}{\Gamma _\phi }/{\Gamma _{{\rm{em}}}} + {\beta _{C,{\rm{em}}}}{\Gamma _C}}T_{\rm em} ,
\end{equation}
and imperfections of the photon emission
\begin{equation} \label{fid_ph_em}
\xi _{{\rm{em}}}^{{\rm{ph}}} =  - {\beta _{{\rm{em}}}}\log \left( {(1 - {e^{ - {\Gamma_{\rm em}T_{\rm em}}}})\frac{{{\Gamma _{{\rm{em}}}}}}{{{\Gamma _{{\rm{em}}}} + {\Gamma _T}}}{p_{{\rm{em}}}}} \right).
\end{equation} 

The above overall scaling matches our qualitative prediction [\cref{quali_unit,quali_emit}] very well. The non-universal coefficients $\{ {{\beta _i}} \}$ depend on the target photonic state and the pulse shape, and are extracted from scaling data shown in \cref{apd_errors}. In our example of cluster state generation, using the pulses shown in \cref{qoc_cq}, we obtain
\begin{equation} \label{scale_beta}
\begin{array}{l}
{\beta _C} = {\rm{2.20}},\quad{\beta _T} = {\rm{1.43,}}\quad{\beta _\phi } = {\rm{0}}.{\rm{92}},\\
{\beta _{\phi ,{\rm em}}} = {\rm{0.51,}}\quad{\beta _{{\rm{C}},{\rm{em}}}} = {\rm{0.47,}}\quad{\beta _{{\rm{em}}}} = {\rm{0.47}},\\
{\beta _0} = {\rm{2}}.{\rm{58}} \times {\rm{1}}{{\rm{0}}^{- 4}},\quad{\beta _\alpha }{\rm{ = 46}}.{\rm{7}}.
\end{array}
\end{equation}
Here ${\beta _0}$ corresponds to the imperfect synthesis of the optimal control pulse, and can typically be made negligible. All other $\{ {{\beta _i}} \}$ are of order $O(1)$, except ${\beta _\alpha }$ which correspond to the effect of transmon double excitations, which is particularly large because it scales with the maximum driving amplitude of the transmon, which can be several times larger than $| \chi  |$ (cf. \cref{qoc_cq}). Note that the decoherence due to transmon double excitations [cf.~\cref{fid_unit}] is still relatively small, thanks to the strong anharmonicity $|\alpha|$ that suppress the transmon double excitations, and one can further reduce this leakage error by including higher levels in the pulse optimization~\cite{Rebentrost2009}.

\subsection{Performance estimation of protocol}
\label{mps_perf}
\begin{figure}[h!]
	\centering
\includegraphics[width=0.48\textwidth]{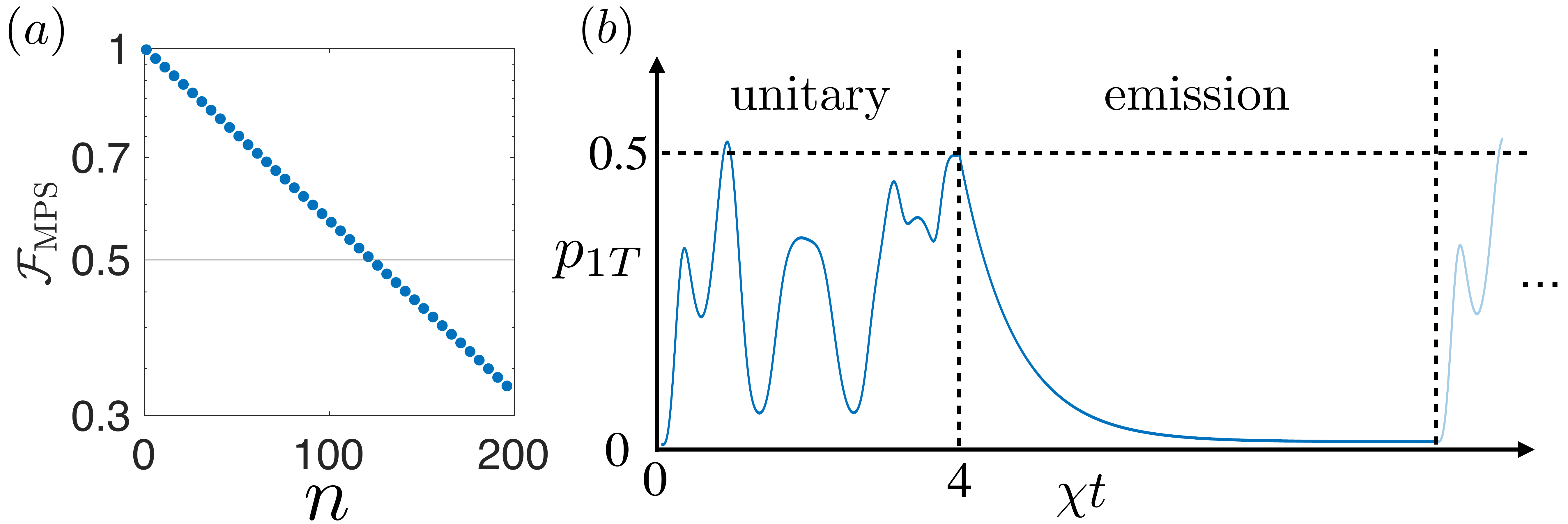}
        \caption{(a) The state fidelity versus the number of photons for the cluster state generated using the pulse sequence in \cref{qoc_cq}, with the state-of-the-art parameters listed in \cref{mps_perf}. The horizontal line denotes ${\cal F} _{\rm MPS}=1/2$, with the corresponding photon number $N_{\rm ph}$ defined as the \textit{entanglement length} of the photon string. (b) Evolution of the transmon excited population $p_{1T}$ during one sequence of cluster state generation. The $p_{1T}$ shows a transient evolution during the unitary operation $U_{[i \neq n]}$ driven by the pulse in \cref{qoc_cq}, and exponentially decays during the photon emission process.}
        \label{pe_plot}
\end{figure}

We estimate the performance of this sequential photon source using current state-of-the-art experimental parameters. Specifically, we use cavity and transmon parameters~\cite{Heeres2017} ${\Gamma _C} = \SI{0.37}{\rm kHz}$, ${\Gamma _T} = \SI{5.88}{\rm kHz}$, ${\Gamma _\phi } =\SI{23.26}{\rm kHz}$, 
$\alpha  = 2\pi  \times - \SI{236}{\rm MHz}$, $\chi  = {\rm{2}}\pi  \times  - \SI{2.194}{\rm MHz}$, and the photon emission parameters~\cite{Besse2020} ${\Gamma _{\rm em}} = 2\pi  \times \SI{1.95}{\rm MHz}$, $p_{\rm em}=1$~\footnote{Here we assume that the realization of our protocol using cavity + transmon system can also be modeled with the photon retrieval efficiency $p_{\rm em}=1$, same as that in Ref.~\cite{Besse2020}. }. We choose the photon emission duration as $T_{\rm em}^{\rm opt} = \log (1 + {\beta _{{\rm{em}}}}{\Gamma _{{\rm{em}}}}/{\beta _{{\rm{C}},{\rm{em}}}}{\Gamma _C})/{\Gamma _{\rm em}}$ to minimize $\xi$ [cf.~\cref{fid_scale}] under the assumption of a fixed photon emission rate ${\Gamma _{\rm em}}$. For the one-dimensional cluster state generated using the pulse sequence in \cref{qoc_cq}, we plot the resulting fidelity ${\cal F}_{\rm MPS}$ in \cref{pe_plot}(a). Defining the \textit{entanglement length} ${N_{\rm ph}} = \log 2/\xi $ (which $\xi$ has been defined in \cref{err_scale}), which is the photon number at which the fidelity drops down to ${\cal F}_{\rm MPS}=1/2$, we obtain ${N_{\rm ph}} \approx 123$,
which would mean an eightfold increase compared to the experimentally demonstrated ${N_{\rm ph}} \approx 15$ in Ref.~\cite{Besse2020}. This improvement partly comes from the fact that we exploit the long transmon lifetime reported in Ref.~\cite{Heeres2017}. If we use the transmon properties reported in \cite{Besse2020}, our protocol gives ${N_{\rm ph}} \approx 47$ (see \cref{beta_scale}), which is still a substantial improvement. 

To understand this improvement, we plot the population $p_{1T}$ of the transmon excited state $| 1 \rangle_T $ during one photon generation round in \cref{pe_plot}(b). We see that $| 1 \rangle_T $ is only transiently populated during the unitary operation $U_{[i \neq n]}$ (driven by the pulse in \cref{qoc_cq}), compared to the protocol in Ref.~\cite{Besse2020} where one always has ${p_{1T}} = 1/2$ during the cluster state generation. This leads to a substantial improvement of the entanglement length obtained in our protocol~\footnote{Since the emitter is always in its ground state at the start of each unitary operation (i.e. after each photon emission), the transmon will only transiently populated during each unitary operation.}. Moreover, it is possible to further reduce $p_{1T}$ by adding a corresponding penalty in the optimal control algorithm (cf. \cref{goat_imple}).

Finally, we estimate the scaling of $N_{\rm ph}$ with the bond dimension $D$ of the desired MPS as (see \cref{mps_bond})
\begin{equation} \label{}
{N_{{\rm{ph}}}} \propto {D^{ - 2}},
\end{equation}
and this scaling mainly comes from the time $T_{\rm MPS}^D = O(D^2)$~\cite{Lee2018a} taken to implement a unitary on the $2D$-dimensional Hilbert space $\dim {\cal H_{\rm src}}=2D$
(cf.~\cref{cq_mps_prot}). This scaling indicates that our system can create MPS of moderate bond dimensions. This already finds many applications and can efficiently capture the ground states of one-dimensional local gapped Hamiltonians \cite{Huang2015a,Dalzell2019,Huang2019a,Schuch2017}. 


\section{Generating rp-PEPS with cQED}
\label{hd_iso}

The previous section shows that the proposed cavity-transmon system can produce high-fidelity one-dimensional photonic MPS. In this section, we demonstrate how to extend this to implement the high-dimensional photonic state generation protocol introduced in Ref.~\cite{zypp}. First, in \cref{arr_set} we introduce the array of coupled sequential photon sources and prove the universality of the Hamiltonian for this system, which allows this system to implement arbitrary local unitary transformations. Then we show in \cref{psg_cq} that, using this setup, one can generate radial plaquette PEPS (rp-PEPS), whose \textit{source point} (cf.~\cref{psg_cq}) sits on a corner of the lattice. After that, in \cref{rp_example} we demonstrate this protocol by discussing the preparation of two-dimensional cluster state, the toric code, and the isometric tensor network states (isoTNS). Finally, we analyze the scaling of the state preparation fidelity for this protocol in \cref{rp_fid}. Here we mainly focus on generating two-dimensional photonic states, however, this protocol readily extends to higher spatial dimensions~\cite{zypp}.

\subsection{Setup: array of sequential photon sources}
\label{arr_set}
\begin{figure}[h!]
	\centering
\includegraphics[width=0.48\textwidth]{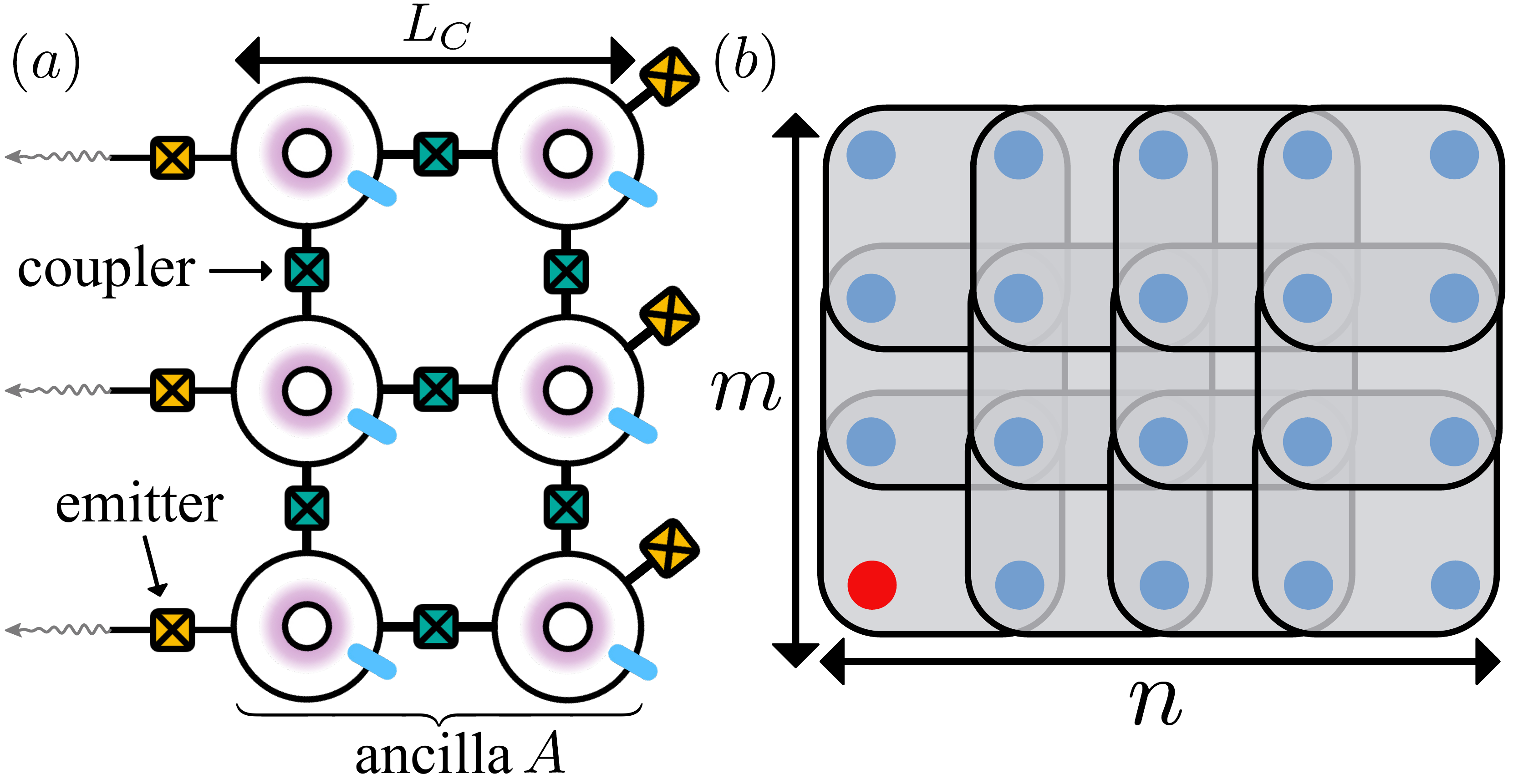}
        \caption{ (a) The setup to generate photonic radial plaquette PEPS (rp-PEPS), which consist a quasi-one-dimensional array of cavity-transmon pairs (the notations follow that in \cref{fig1}a). In each row there is a transmon emitter that can emit photons. The neighboring cavities are connected by Y-shape couplers [the green (dark gray) boxes]. (b) rp-PEPS with open boundary conditions are produced by sequential applying unitaries on overlapping regions. Here the \textit{source point}~\cite{zypp} of the state is denoted by the red dot.}
                \label{psg_gen_set}
\end{figure}

Let us consider a natural generalization of the setup in \cref{cq_mps_set} to a quasi-one-dimensional array of $L_c\times m$ cavity-transmon pairs, and use the first $D'$ Fock states of each cavity. The $L_c$ cavities in each row form an ancilla $A$ of dimension $D = D{'^{{L_c}}}$. While $L_c=1$ is the most straightforward choice, sometimes it may be beneficial to choose $L_c>1$, as we comment on in \cref{rp_fid}. Moreover, in each row there is one \textit{emitter} ${\{ {{T_j}} \}_{j = 1,...,m}}$ (see \cref{psg_gen_set}a), while other transmon qubits are used to provide universal control to the corresponding cavities~\footnote{Here we do not include the transmons in the ancilla space due to their relatively short coherence time compared to the photons in the cavities. However, in principle, one can include the transmons into the ancilla as well to further increase its Hilbert space dimension.}.

We couple each neighboring pair of photon sources by a coupler that interacts with both cavities \cite{Wang2016} (shown as green boxes in \cref{psg_gen_set}a). By driving the coupler with two-tone pumps, the four-wave mixing process of the coupler reduces to the following bilinear interaction \cite{Gao2018}
\begin{equation} \label{bilin_ham}
 H_{{\rm{int}}}^{ij}(t) = {g_{ij}}(t)( {{e^{i{\varphi _{ij}(t)}}}a_i^\dag {{a}_j}  + {\rm H.c.}} ).
\end{equation}
Here $a_i^\dag$ denote the creation operator of the cavity for $i$-th photon source. The coupling strength $g_{ij}( t )$ and phase $\varphi_{ij} ( t )$ can be controlled through drives. Let us denote the set of vertices of the array of sources in \cref{psg_gen_set}(a) as $\cal V$, where each vertex $v_{ij}$ connects the cavities of the $i$-th and $j$-th cavity-transmon pairs. One can thus write down the Hamiltonian of this system as
\begin{equation} \label{peps_ham_tot}
{H_{{\rm{array}}}}(t) = \sum\limits_{i = 1}^{{L_c} \times m} {H_{{\rm{src}}}^i(t)}  + \sum\limits_{{v_{ij}} \in {\cal V}} {H_{{\rm{int}}}^{ij}} (t),
\end{equation}
where $H_{\rm src}^i$ is the Hamiltonian for the $i$-th source, containing the terms in \cref{mps_ham_dri,mps_ham_ctl}. Equipped with the universal control of each source and the bilinear couplings, we prove in \cref{apd_univ} that $H_{\rm{array}}$ provides universal control. Given this, we can assume that one can implement arbitrary local unitary operations one this system.

In sum, this system can be represented by a one-dimensional array of $D{'^{{L_C}}}$-level ancillas (labeled by ${\{ {{A_j}} \}_{j = 1,...,m}}$) coupled to transmon emitters (labeled by ${\{ {{T_j}} \}_{j = 1,...,m}}$), illustrated in \cref{psg_gen}. Note that one can let more transmons in each row to emit photons, which effectively increase the dimension of the transmon emitters. Finally, in the next sections we will treat each ancilla effectively as $L_p-1$ qubits by choosing $D'$ and $L_c$ such that 
\begin{equation} \label{anci_dim}
D{'^{{L_C}}} \ge {2^{{L_p} - 1}}. 	
\end{equation}

\subsection{Preparation of rp-PEPS}
\label{psg_cq}

Ref.~\cite{zypp} introduces a generic protocol to produce rp-PEPS on flying qubits. rp-PEPS are states prepared by sequentially applying unitaries on plaquettes of size $L_p \times L_p$ ($L_p \ll n,m$) in a radial fashion [see \cref{psg_gen_set}(b) for an example]. They possess long-range correlations and area-law entanglement, and photonic rp-PEPS can be efficiently prepared with the circuit depth \cref{circ_dep}.

 \begin{figure}[h!]
	\centering
\includegraphics[width=0.48\textwidth]{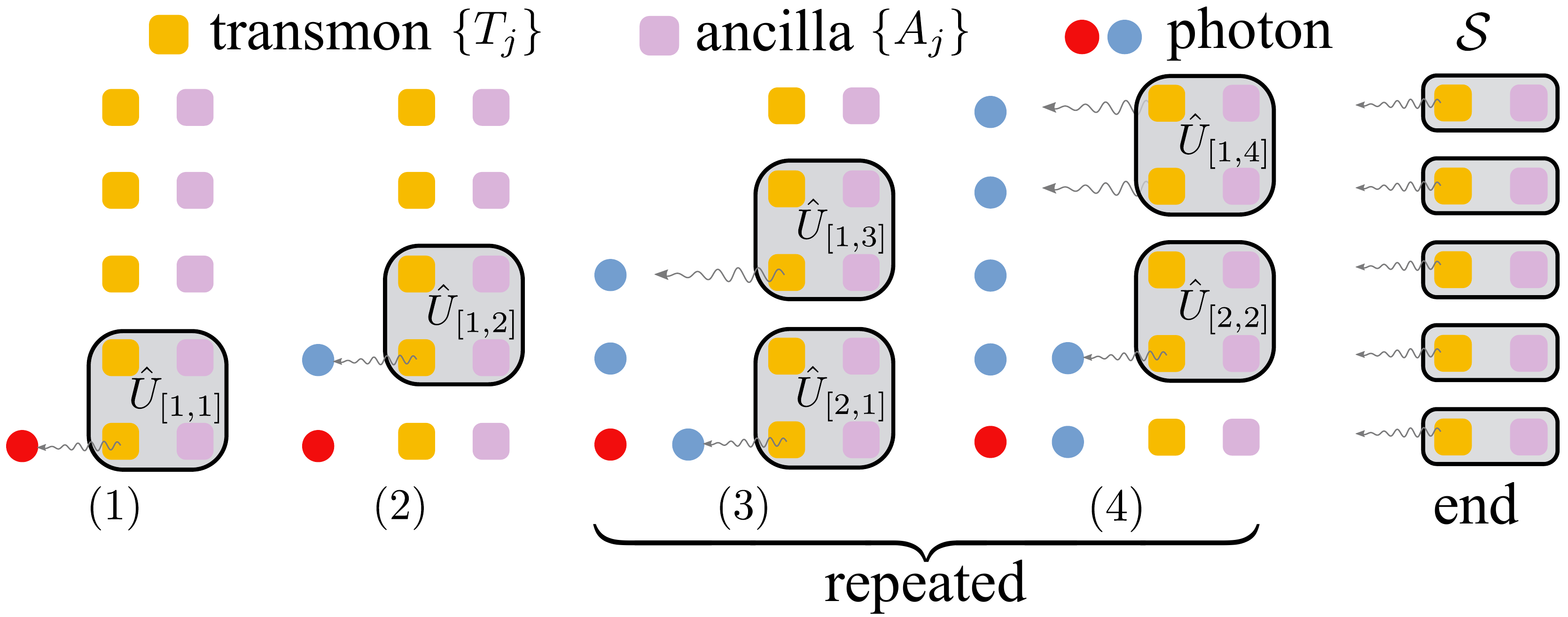}
        \caption{The preparation of photonic rp-PEPS ($L_p=2$ here). In the preparation of the $(i,j)$-th site, we apply a unitary ${\hat U_{[ {i,j} ]}}$ followed by a photon emission of the transmon $T_j$. After the initial steps (1) and (2), steps (3) and (4) will be repeated. At the end of the protocol, we swap the excitations on the ancillas to the emitters and then convert them to photons, denoted as $\cal S$.}
                \label{psg_gen}
\end{figure}

 The generation procedure of rp-PEPS is shown in \cref{psg_gen}.  We start from an initial state where all the cavities and the transmons are in their ground states $| {{\varphi _0}} \rangle  = | {{0_{\{ {{A_j}} \}}}} \rangle  \otimes | {{0_{\{ {{T_j}} \}}}} \rangle $,
and sequentially apply unitaries $\{ {{{\hat U}_{\left[ {i,j} \right]}}} \}$. The unitary $U_{[i,j]}$ is applied in the $i$-th layer on the ancillas $\{ {{A_j},...,{A_{j + {L_p} - 1}}} \}$ and transmons $\{ {{T_j},...,{T_{j + {L_p} - 1}}} \}$. As each ancilla can be viewed effectively as $L_p-1$ qubits~\cref{anci_dim}, the unitary equivalently acts on a plaquette of qubits of size $L_p \times L_p$. After each unitary, we trigger the photon emission from the transmon $T_j$ (denoted as isometry $M^{j}_{{\rm{ph}}}$ [cf.~\cref{ap_map}]). Note that after the last unitary per column, the last emission process per column $M^{m-L_p+1}_{{\rm{ph}}}$ (see step 4 of \cref{psg_gen}) convert excitations of transmons $\{ {{T_{m - {L_p} + 1}},...,{T_m}} \}$ to multiple photonic qubits at the same time. Repeatedly applying this procedure following the order shown in \cref{psg_gen} (also see the below \cref{pP_ph}), and in the end emitting the remaining excitations in the ancillas (an operation collectively denoted as $\cal S$), we generate the desired two-dimensional photonic state~\cite{zypp}
\begin{equation} \label{pP_ph}
|{\psi _{{\rm{rp}}}}\rangle = \langle \varphi _{0 }| {\cal S}\prod\limits_{i = 1}^{n-L_p+1} {\prod\limits	_{j = 1}^{m-L_p+1} {(M_{{\rm{ph}}}^j{{\hat U}_{[i,j]}})} |\varphi_{0}\rangle }.
\end{equation}

Given the universal control of \cref{peps_ham_tot}, this protocol can produce arbitrary states of form \cref{pP_ph} (schematically shown in \cref{psg_gen_set}b), which are two-dimensional rp-PEPS of plaquette size $L_p$ with open boundary condition, with their \textit{source point} located at the first photon being created~\footnote{The source point of rp-PEPS is precisely defined as the location of the first plaquette unitary~\cite{zypp}.}. Also note that, in the above protocol, photon emissions reset the transmon. This allows one to efficiently reuse them, and this parallelize the preparation procedure (shown in steps 3 and 4 of \cref{psg_gen}), such that the circuit depth for preparing rp-PEPS of plaquette length $L_p$ on a $n \times m$ lattice asymptotically scales as \cref{circ_dep}~\cite{zypp}. This system further allows increasing the plaquette size by increasing the number of cavity-transmon pairs $L_C$ or using more modes $D'$ in the cavity.

Finally, we note that by replacing the photon emissions with qubit measurements, this state generation protocol naturally becomes a qubit-efficient quantum variational scheme~\cite{Huggins2019,Liu2019}.

\subsection{Examples}
 \label{rp_example}
\subsubsection{Two-dimensional cluster state}
\label{cls_rp}

Consider a two-dimensional square lattice, with the position vector of each site denoted as $\vec a \equiv \left( {i,j} \right)$. The two-dimensional cluster state ${| {\rm Cl} \rangle _{{\rm{2D}}}}$ is prepared starting from a product state of all qubits in $\left|  +  \right\rangle$ by applying control-Z (CZ) gates on each nearest-neighbour pairs of qubits~\cite{Briegel2001}
\begin{equation} \label{cls2d_eq}
{| {\rm Cl} \rangle _{{\rm{2D}}}} =  {\prod\limits_{\vec b{\textrm{ adjacent }}\vec a} {{\rm{C}}{{\rm{Z}}_{\vec a\vec b}}} }  \textrm{ } \mathop  \bigotimes \limits_{\left\{ {\vec a} \right\}} \left|  +  \right\rangle.
\end{equation}

\begin{figure}[h!]
	\centering
\includegraphics[width=0.48\textwidth]{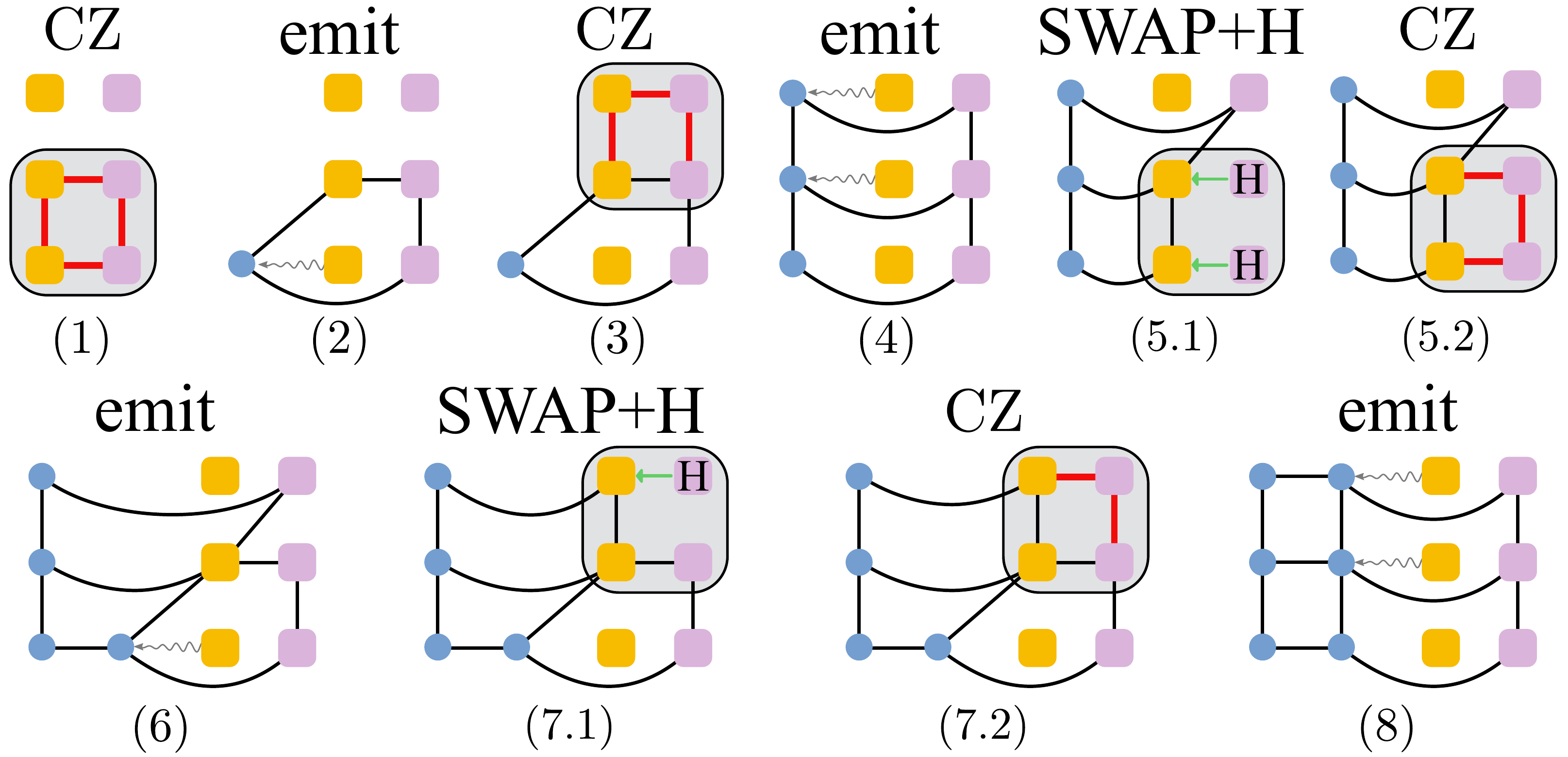}
        \caption{Preparing two-dimensional cluster state ${| {\rm Cl} \rangle _{{\rm{2D}}}}$ following the rp-PEPS preparation protocol (cf.~\cref{psg_gen}, and we keep the same notation as there). Here both the ancillas and transmons have two levels (qubits), initialized in the state $|  +  \rangle $. The lines connecting two qubits mean that they have been acted on by a control-Z (CZ) gate in a previous step, whereas the thick red (gray) lines denotes that CZ gates are being applied in that step. 
        The steps from 1 to 4 prepare the first column of photons. To create the next column of photons, in each unitary we first swap certain ancilla states to the corresponding transmons in the same row [illustrated by green (gray) arrows], then apply Hadamard gates to set the ancilla states to $\left|  +  \right\rangle $ (see 5.1 and 7.1), and finally before apply CZ gates, which prepares the second column (see step 8). By repeating steps 5.1 to 8, one can prepare ${| {\rm Cl} \rangle _{{\rm{2D}}}}$ of arbitrary size.}
        \label{cls2d}
\end{figure}

This state is an rp-PEPS of plaquette size $L_p=2$, thus it can be prepared by the protocol in \cref{psg_cq}, where each ancilla consists of a qubit (this means that we use $L_c=1$ cavity with $D'=2$ Fock basis). The preparation procedure is shown in \cref{cls2d}, where the corresponding plaquette unitaries $\{ {{U_{\left[ {i,j} \right]}}} \}$ in \cref{pP_ph} are formed by CZ, SWAP, and Hadamard gates.

One can easily extend the size of the state horizontally by repeatedly applying the steps 5.1-8 in \cref{cls2d}, and vertically by fabricating a longer chain of ancilla-transmon pairs. Moreover, omitting some photon emissions, one can create arbitrary graph states of local connectivities in this way~\cite{Russo2019}.

To prepare ${| {\rm Cl} \rangle _{{\rm{2D}}}}$ of size $n\times m$, the depth for this circuit in terms of plaquette unitaries is ${\cal T} \approx 2 n + m$ [cf.~\cref{circ_dep}]. We also point out that this can be further improved to ${\cal T_{\rm cl}} \approx 2 n$, since the CZ gates commute with each other. As an example, steps 1-4 in \cref{cls2d} can be combined such that we apply CZ gates on all adjacent pairs in the ancilla-transmon array in parallel (which are contained in two layers of plaquette unitaries), then emit one column of photons. The CZ gates between two cavities can be realized by combining single-qubit rotations and a CNOT gate, which has been experimentally implemented in Ref.~\cite{Rosenblum2018a}.

The above further parallization of the circuit show a generic feature of photonic rp-PEPS [\cref{pP_ph}]: if the sequential product of unitaries $\hat U_{{\rm{col}}}^i = \prod\nolimits_{j = 1}^{m - {L_p} + 1} {{{\hat U}_{\left[ {i,j} \right]}}} $ for preparing each column of photons can be parallelized as a circuit of depth $O(1)$, one can directly implement $\hat U_{{\rm{col}}}^i$ followed by the photon emission of all transmon emitters $\mathop  \otimes \limits_j M_{{\rm{ph}}}^j$ to prepare the $i$-th column of the rp-PEPS. This also applies to the toric code below.

\subsubsection{The toric code}
\label{tc2d}

\begin{figure}[h!]
	\centering
	\includegraphics[width=0.48\textwidth]{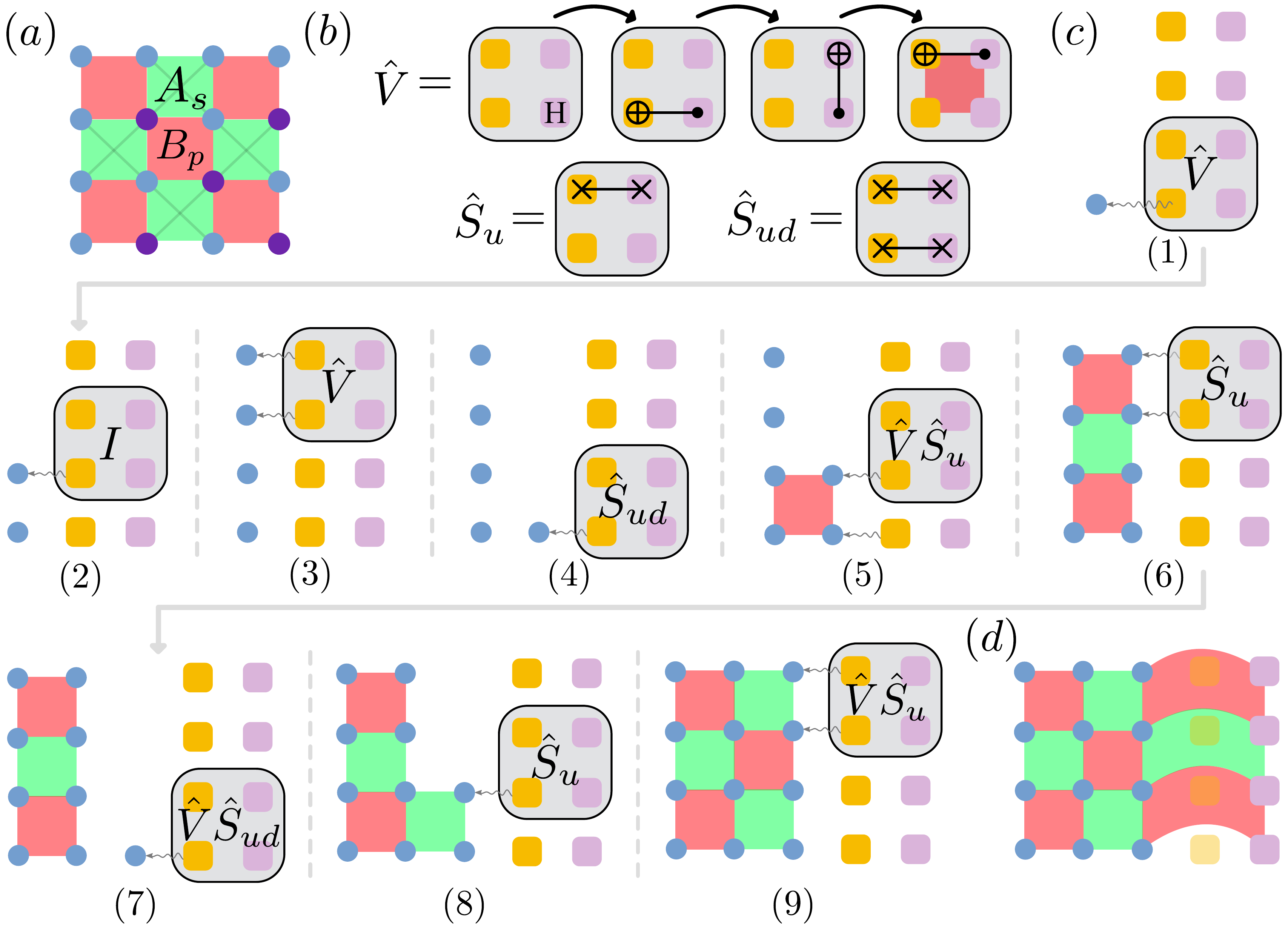}
        \caption{Preparation of photonic toric code $|{\rm TC}\rangle$, inspired by Ref.~\cite{satzinger2021}. Throughout this figure, we keep the same notation as that in \cref{psg_gen}. The support of the star operators $A_s$ is shown in green (light gray), and that of the plaquette operators $B_p$ in red (dark gray).
        (a) The toric code is the unique $+1$ eigenstate of all star and plaquette operators $A_s$ and $B_p$. Each plaquette has a `representative qubit', denoted by a purple dot.
        (b) We group the Hadamard and CNOT gates that will be applied to each plaquette as $\hat V$. The SWAP gates are denoted as ${\hat S_u}$ and ${\hat S_{ud}}$. (c) We alternatively apply $\hat V$ or identity gate $I$ in each step (dashed lines separate different steps), followed by a photon emission (for example, from steps 1 to 3). Also after each photon emission, one needs to swap the corresponding ancilla state to the transmon, thus in steps 4-9 the unitaries are generally the product of swap gates (${\hat S_u}$ or ${\hat S_{ud}}$) with $\hat V$ or $I$. (d) After step 9 of the panel (c), one obtains $|{\rm TC}\rangle$ of size $4 \times 4$ with the same geometry as that in panel (a), consisting of three columns of photonic qubits and the column of ancilla qubits. By further repeating the steps from 4-9 in panel (c) and using more ancilla-transmon pairs, one can generate $|{\rm TC}\rangle$ of arbitrary size.}
        \label{tc_fig}
\end{figure}

The toric code \cite{Dennis2002,Kitaev2003} is an example of a string-net state, and finds important applications in quantum error correction. It is defined as the simultaneous $+1$ eigenstate of all star operators $A_{s}=\prod_{i \in s} Z_{i}$ (green boxes in \cref{tc_fig}a) and plaquette operators $B_{p}=\prod_{j \in p} X_{j}$ (red boxes in \cref{tc_fig}a), i.e. $A_s | {{\rm{TC}}} \rangle = B_p | {{\rm{TC}}} \rangle = +1| {{\rm{TC}}} \rangle$. Here $\{Z_i\}$ and $\{X_j\}$ are Pauli operators.

The toric code has recently been prepared on a stationary lattice, with the following procedure~\cite{satzinger2021}
\begin{enumerate}
	\item Initialize the whole lattice in the state $\mathop  \bigotimes \limits_{\left\{ {\vec a} \right\}} \left| 0 \right\rangle $, where all $\langle {{A_s}} \rangle  = 1$ and $\langle {{B_p}} \rangle  = 0$.
	\item Choosing a qubit for each plaquette as the representative qubit (an example choice is denoted by purple dots in \cref{tc_fig}a), and apply Hadamard gate on it.
	\item Within each plaquette, sequentially apply CNOTs with the representative qubit as the control and other qubits as targets, with an ordering such that the representative qubits are not changed until the CNOTs in their plaquette have been applied [cf.~\cref{tc_fig}(b)].
\end{enumerate}

Inspired by this procedure, one can prepare $|{\rm TC}\rangle$ as a photonic rp-PEPS of $L_p=2$ using the protocol in \cref{psg_cq}, with each ancilla consisting of a single qubit. To ease the notation, in \cref{tc_fig}(b) we group the gates in steps 2-3 of the above procedures that act on each plaquette as $\hat V$, and the gates used to swap the ancilla and transmon state as ${\hat S_u}$ and ${\hat S_{ud}}$.

The preparation circuit is shown in \cref{tc_fig}(c), where the plaquette unitaries are alternatively formed by $\hat V$ or the identity gate $\hat I$, and their product with swap gates (${\hat S_u}$ or ${\hat S_{ud}}$). The $\hat V$ implement desired operations for each toric code plaquette region~\cite{satzinger2021}, while the photon emission and swap gates will `push' the produced entangled states toward the photon lattice (denoted by blue circles). After all operations in \cref{tc_fig}(c), one obtains $|{\rm TC}\rangle$ of size $4\times 4$ on three columns of photonic qubits and the ancilla qubits, shown in \cref{tc_fig}(d). By further repeating the steps from 4-9 in \cref{tc_fig}(c) and using more ancilla-transmon pairs, one can generate $|{\rm TC}\rangle$ of arbitrary size. We also point out that, by further parallelizing the unitaries~\cite{satzinger2021}, one can apply the same procedure as described in the two-dimensional cluster state generation protocol [cf.~\cref{cls_rp}] to obtain a circuit depth ${{\cal T}_{\rm TC}} \approx  {2 n} $ in terms of plaquette unitaries for generating $\left| {\rm TC} \right\rangle $ of size $n \times m$.

The schemes presented here for cluster state generation and toric code generation are directly derived from their circuit generation on stationary lattices. This idea allows one to obtain photonic state generation circuits by utilizing existing circuits on stationary lattices. For example, one can generate string-net states by extending the protocol in \cref{tc2d}, using similar circuits as in Ref.~\cite{yj_circ}.

\subsubsection{Isometric tensor network states}
\label{iso_ex}
The rp-PEPS contain the isometric tensor network states~\cite{Zaletel2020} (isoTNS) as a subclass~\cite{zypp}. An isoTNS is parametrized by its bond dimension $D$~\cite{Zaletel2020} which bounds the entanglement entropy of the state, and its physical dimension $d$, which specifies the Hilbert space dimension of each site.

 \begin{figure}[h!]
	\centering
	\includegraphics[width=0.48\textwidth]{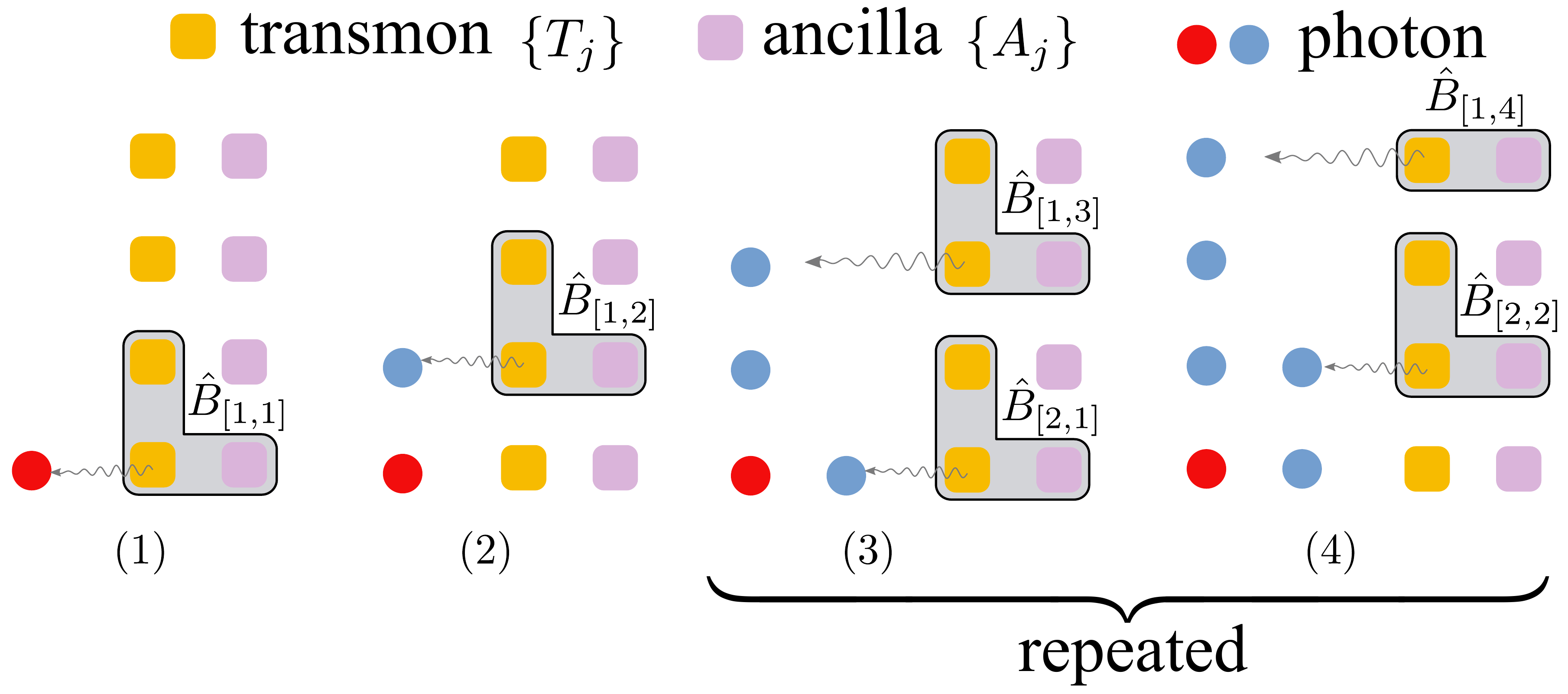}
        \caption{Preparation of isoTNS with bond dimension $D=2$ and physical dimenison $d=2$. In the preparation of the $(i,j)$-th site, we apply a unitary ${\hat B_{[ {i,j} ]}}$
         connecting the cavity $C_j$ and the transmons $T_j,T_{j+1}$, followed by a photon emission of the transmon $T_j$. After the initial steps (1 and 2), steps 3 and 4 will be repeated to build up the desired 2D isoTNS. The red (dark gray) dot denotes the orthogonality center of the isoTNS.}
        \label{isop_tori}
\end{figure}

As the protocol in \cref{psg_cq} can prepare rp-PEPS with the source point in the corner of the lattice, one can prepare the subclass of isoTNS whose \textit{orthogonality center} (see details in \cref{iso_psg}) is in this corner. To do so, one has to require the unitaries to have an `L'-shape~\cite{zypp}, as shown in \cref{isop_tori} for the case of $D=2$ and $d=2$. Here each unitary $\hat B_{[i,j]}$ acts on the ancilla $A_j$ and transmons $\{ {{T_j},...,{T_{j + {L_p} - 1}}} \}$ to produce isoTNS of bond dimension $D \le 2^{L_p-1}$. Increasing isoTNS bond dimension corresponds to increasing the arm length of the `L'-shaped unitary. In the end of the protocol, we can disentangle the ancilla from the photonic state being produced~\cite{zypp}. We provide more details on isoTNS in \cref{iso_psg}.

The subclass of isoTNS whose orthogonality center is in the corner of the lattice already contains all graph states of local connectivities~\cite{Russo2019} and all string-net states~\cite{Soejima2020}, and thus the two-dimensional cluster state and toric code discussed previously can be also created in this way. However, we point out that, the circuits we presented in \cref{cls_rp,tc2d} are more efficient than the circuit derived from their isoTNS representation. For example, the isoTNS representation of arbitrary $\mathbb{Z}_\lambda$ toric code~\cite{Bullock2007} is shown in \cref{tc_iso}, where the physical dimension of the tensor is $\lambda^4$. Thus for the qubit ($\lambda=2$) toric code ($| {{\rm{TC}}} \rangle$ discussed in \cref{tc2d}), this isoTNS generation scheme requires each transmon emitter to be 16-dimensional, for which we need to couple multiple transmons to the same ancilla, thus is more complex than the scheme shown in \cref{tc2d}. The existence of isoTNS representation of $\mathbb{Z}_\lambda$ toric code for arbitrary $\lambda$ also implies that our scheme can create photonic qudit toric code, and such states show certain advantages as quantum error-correcting codes compared to the qubit toric code~\cite{Duclos-Cianci2013,Watson2015}.

Finally, isoTNS is a class of states that can be efficiently prepared as shown here (also see Ref.~\cite{zypp}), and serve as an ansatz for classical variational algorithms~\cite{Zaletel2020}. Combining these two features may allow one to implement interesting protocols such as variational quantum metrology~\cite{Jarzyna2013,Koczor2020,Chabuda2020}.

\subsection{Scaling of rp-PEPS generation fidelity}
\label{rp_fid}
In the presence of imperfections, the rp-PEPS generation protocol [cf.~\cref{psg_cq}] will yield a mixed photonic state ${\rho _{\rm rp}}$. Here we provide a qualitative estimation of the fidelity ${{\cal F}_{\rm rp}} = \langle {{\psi _{\rm rp}}} |{\rho _{\rm ph}}| {{\psi _{\rm rp}}} \rangle $ of the rp-PEPS generation protocol. Moreover, one can calculate the fidelity exactly by extending the MPDO approach in \cref{mpdo_fid}.

Let us consider the array shown in \cref{psg_gen}(b) with $m$ rows, where each row consists of $L_C$ cavity-transmon pairs, with each cavity and transmon having the decoherence channels described in \cref{model_err}. Since we use the first $D'$ Fock states of the cavity, we take the worst-case estimation of the cavity decay rate ${\Gamma' _C} = (D'-1)\Gamma_C$.

To prepare a generic rp-PEPS of size $n \times m$ and plaquette length $L_p (L_p \ll n,m)$, when preparing each photon we need to apply a unitary acting on $L_p^2$ qubits, thus $\dim {{{\hat U}_{\left[ {i,j} \right]}}} \approx 2^{L_p^2}$. We estimate the time $T_{\rm rp}$ of implementing a generic unitary using optimal control methods as $T_{\rm rp} \sim {( {\dim {{\hat U}_{\left[ {i,j} \right]}}})^2} = O( {{4^{L_p^2}}} )$~\cite{Lee2018a}. Note that the numerical cost of the pulse optimization also increases exponentially with $L_p^2$. Since both the initial and the final state of the sources are the ground state $\left| {{\varphi _0}} \right\rangle $ [cf.~\cref{pP_ph}], in the preparation procedure [cf.~\cref{psg_gen}], each source remains excited for a time $L_p n T_{\rm rp}$. Moreover, there are in total around $nm$ photon emissions. Thus by applying the same argument as in the MPS generation protocol (cf.~\cref{model_err}), one can estimate the scaling of the fidelity as
\begin{equation} \label{iso_fid_eq}
{{\cal F}_{{\rm{rp}}}}\sim\exp \left[ { - {\xi '}^{D'}_{{\rm{unit}}} \cdot m{L_c} \cdot {L_p}n{T_{\rm rp}} - ( {{\xi '}_{{\rm{em}}}^{{{\rm src},D'}} + {\xi '}_{{\rm{em}}}^{{\rm{ph}}}} ) \cdot nm} \right],
\end{equation}
with ${\xi '}^{D'}_{{\rm{unit}}}$, ${\xi'} _{\rm em}^{{{\rm src},D'}}$ and ${\xi'} _{\rm em}^{\rm ph}$ of the same form as that in \cref{fid_unit,fid_src_em,fid_ph_em}, but with cavity decay rate $\Gamma_C$ replaced by $\Gamma'_C$ and different non-universal constants $\{ {{\beta' _i}} \}$. 

Given $L_p$, the number of cavities in each ancilla $L_C$ is ${L_C} = \left\lceil {( {{L_p} - 1} )/{{\log }_2}D'} \right\rceil $. We can thus write overall scaling [\cref{iso_fid_eq}] as ${\cal F}_{\rm rp}\sim\exp ( { - {\xi^{L_p,D'} _{\rm rp}} \cdot nm} )$, with the error rate of generating each photon ${\xi^{L_p,D'} _{\rm rp}}$ as
\begin{widetext}
\begin{equation} \label{rp_err_scale}
\xi _{{\rm{rp}}}^{{L_p},D'} \approx   {4^{L_p^2}} ( {{L_p} - 1} ){L_p}\cdot{\xi '}_{\rm unit}^{D'}/{\log _2}D' +  {{\xi'} _{\rm em}^{{{\rm src},D'}} + {\xi'} _{\rm em}^{\rm ph}} .
\end{equation}
\end{widetext}

From the above analysis, we see that using more cavities $L_c>1$ reduces the maximum number of excitations in each cavity $D' = \left\lceil {2^{(L_p-1)/L_c}} \right\rceil$ significantly, which leads to a corresponding reduction in the error due to cavity decay $\Gamma'_C=(D'-1)\Gamma_C$. In addition, it is experimentally easier to control cavities with small Hilbert space than controlling a single cavity with a large Hilbert space. However, the fidelity of the inter-cavity connection is generally lower than the single cavity operation~\cite{Gao2018}. So depending on the desired plaquette length $L_p$ of the rp-PEPS, one needs to choose appropriate $L_c$ and $D'$ to get the highest possible fidelity. A precise determination of the optimal $L_c$ and $D'$ further depends on the specific target state, the details of the unitary operations, as well as the errors in the inter-cavity operations~\cite{Gao2018, Zhang2019e}.


We expect an experimental realization of the simplest $L_c=1$ protocol could be a good first step to further understand the performance of the rp-PEPS protocol. In the end, an experiment aiming at large plaquette sizes may have to be optimized regarding the total number of cavities used, but this is beyond the scope of the present paper.





\section{Conclusion}
\label{conclu}
We propose a physical platform and a protocol to sequentially generate microwave photonic tensor network states with moderately high bond dimensions based on a dispersively coupled cavity-transmon system. The good coherence properties of microwave cavities lead to favorable scaling of the photon number for the MPS, in particular, we show this platform can potentially create a one-dimensional cluster state of over a hundred photons deterministically with current technology. The good connectivity makes this platform a promising candidate for generating a large class of high-dimensional rp-PEPS, and we show how to create a two-dimensional cluster state, the toric code, and isoTNS as examples. Our work thus serves as systematic guidance for sequential photon generation experiments in circuit QED platforms, and can naturally be applied to other dispersively coupled qubit-oscillator systems.

Our work can be extended in many ways. First, there are plenty of ideas to further reduce the imperfections during the protocol by applying error-correction techniques \cite{Ofek2016,Campagne-Ibarcq2020}, applying error-transparent gates or path-independent gates \cite{Ma2019a} on the bosonic modes, or applying open system optimal control techniques \cite{Schulte-Herbruggen2011,Machnes2018,Abdelhafez2019}. Second, the ability to generate strongly correlated photonic tensor network states opens the door of developing quantum information processing protocols that go beyond MPS~\cite{Jarzyna2013, Chabuda2020, Koczor2020}. One can further simultaneously use coupled arrays of emitters and non-Markovian feedback approaches~\cite{Pichler2017, Dhand2018, Xu2018,Wan2020,Zhan2020,Shi2021,Bombin2021} to reduce the component overhead of the system and possibly generate a larger class of photonic states. 

\begin{acknowledgements}
We thank Yujie Liu and Alejandro Gonz\'alez-Tudela for their insightful discussions.	We acknowledge funding from ERC Advanced Grant QUENOCOBA under the EU Horizon 2020 program (Grant Agreement No. 742102) and the European Union's Horizon 2020 research and innovation program under Grant No. 899354 (FET Open SuperQuLAN).
\end{acknowledgements}

\appendix

\section{MPS generation with cQED using Quantum Optimal Control Approach}
\label{goat_imple}
In this section, we introduce our quantum optimal control (QOC) approach (similar to that used in \cite{zyryd}) to implement unitaries on this setup.

We aim to find the driving amplitude ${{\varepsilon _C}( t ),{\varepsilon _T}( t )}$ for the control ${H_{\rm drive}}( t )$ [\cref{mps_ham_ctl}] that implements the desired unitary operations $U_{[i]}$ [\cref{u_act}] in ${{\cal H}_D} \otimes {{\cal H}_T}$.

Given the target MPS to be prepared [\cref{mps}], one can construct a series of isometries $\{ {{{\hat A}_{[ i ]}}} \}$ that needs to be implemented on the system~\cite{Schon2005}. 
For generating the $n$-photon cluster state [\cref{cls_form}], this construction gives two kinds of isometries ${\hat A_{[ {i \ne n} ]}}$ and ${\hat A_{[ {n} ]}}$, as well as the ancilla initial state ${| {{\varphi _I}} \rangle _C}$ needed in the protocol [cf.~\cref{mps}]:
\begin{equation} \label{iso_form}
\begin{array}{*{20}{l}}
{{{\hat A}_{[i \ne n]}} = \left( \begin{array}{l}
V_{\left[ i \right]}^0\\
V_{\left[ i \right]}^1
\end{array} \right) = \frac{1}{{\sqrt 2 }}\left( {\begin{array}{*{20}{c}}
1&0\\
1&0\\
0&1\\
0&{ - 1}
\end{array}} \right),\quad {{\hat A}_{[n]}} = \left( {\begin{array}{*{20}{c}}
1&0\\
0&0\\
0&1\\
0&0
\end{array}} \right),}\\
{|{\varphi _I}{\rangle _C} = \frac{1}{{\sqrt 2 }}{{(|0\rangle }_C} + |1{\rangle _C}).}
\end{array}
\end{equation}
Note that here ${{\hat A}_{[ {n} ]}}$ is different from all other ${{\hat A}_{[i \ne n]}}$, as it contains an additional operation to disentangle the cavity from the photons.


We can embed above $\{\hat A_{[i]}\}$ into ${U_{[i]}}$ to realize \cref{u_act} [see below \cref{utarg}]. In numerical calculations we keep the first $N_{C}>D$ Fock states in ${\mathcal H}_C$. Thus $\dim {{\cal H}_{C}} = {N_C}$ in our numerical calculation. One can then write $U_{[i]}$ as
\begin{equation} \label{utarg}
{{U}_{[i]}} = \left( {\begin{array}{*{20}{c}}
{{{\hat A}_{[i]}}}&{{B_1}}\\
O&{{B_2}}
\end{array}} \right),
\end{equation}
with its basis vector permuted as
\begin{equation}
{\rm{Base}}(U_{[i]}) \equiv [ {{\rm{Base}}( {{{\cal H}_D} \otimes {{\cal H}_T}} ),{\rm{others}}} ].
\end{equation}
The $O$ is a zero matrix, which physically means that ${U_{[i]}}$ does not cause the population to leak out of ${{\cal H}_D} \otimes {{\cal H}_T}$. The parts $B_1$ and $B_2$ are arbitrary, as long as ${U_{[i]}}$ is a unitary of dimension $2N_{C}$. One needs two kinds of unitaries. 
Each application of ${U_{[ {i \ne n} ]}}$ followed by the photon emission adds one site to the cluster state. The last unitary ${U_{[ {n} ]}}$ followed by a photon emission disentangles the source from the photonic MPS.

To apply quantum optimal control (QOC) to our cQED platform with Hamiltonian $H_{\rm src}$ [\cref{mps_ham_dri,mps_ham_ctl}], we go to the rotating frame to remove energy terms of the transmon and the cavity mode in $H_{\rm src}$, getting
\begin{equation} \label{qoc_h}
{H'_{{\rm{src}}}}(t) = \chi {\sigma _{11}}{a^\dag }a + [ {{\varepsilon _C}(t)a + {\varepsilon _T}(t){\sigma _{01}} + {\rm{H.c.}}}],
\end{equation}
and readily apply the QOC algorithm in Ref.~\cite{zyryd} with the control Hamiltonian ${{H'_{{\rm{src}}}}}( t )$ to find the pulse sequences to implement desired $U_{[i]}$. The pulse sequence to implement ${U_{[ {i \ne n} ]}}$ and ${U_{[ {n} ]}}$ for the cluster state generation are shown in \cref{qoc_cq}, where we choose $N_C=5$.

\section{The Matrix Product Density Operator (MPDO) approach to compute state fidelity}
\label{mpdo_fid}

In this section we recall the MPDO approach \cite{zyryd} to compute the fidelity ${{\cal F}_{{\rm{\rm MPS}}}}$ of the photonic state. We can rewrite the master equation \cref{ext_me} in a vectorized form
\begin{equation}\label{rho_v}
\dfrac{{d{{\vec \rho }_{{\rm{\rm src}}}}( t )}}{{dt}} = {\cal L}( t ){\vec \rho _{{\rm{\rm src}}}}( t ),\qquad {\vec \rho _{{\rm{\rm src}}}} = \sum\limits_{a,b = 1}^{{N_h}} {{\rho _{ab}}} | {a \otimes \bar b} \rangle ,
\end{equation}
Here $\mathcal{L}( t )$ is the Liouville operator, $| a \rangle $ is a basis element in ${\cal H}_{\rm src}$ and $| \bar a \rangle$ represent its complex conjugate. The solution of \cref{rho_v} is 
\begin{equation} \label{lv_sol}
{\vec \rho _{\rm{\rm src}}}( T ) = \mathcal{T}\{ {{e^{\int_0^T {\mathcal{L}( t )dt} }}} \}{\vec \rho _{\rm{\rm src}}}( 0 ) = {W_{\cal L}}{\vec \rho _{\rm{\rm src}}}( 0 ).
\end{equation}
The photon emission process can be described by a process map
\begin{equation} \label{wp_map}
{W_{\rm ph}} = \sum\limits_{i,j = 0}^{1} {\sum\limits_{a, b,c,d = 1}^{N_h} {W_{P,abcd}^{ij}} } | {c,\bar d,i,\bar j} \rangle | {a, \bar b} \rangle,
\end{equation}
which maps ${\vec \rho _{{\rm{\rm src}}}}$ with vectorized basis $| {a,\bar b} \rangle $ to a system-photon joint density matrix with vectorized basis $| {c,\bar d,i,\bar j} \rangle  = | {c,\bar d} \rangle  \otimes {| {i,\bar j} \rangle _{\mathrm{\rm ph}}}$. Thus each photon generation round results in a map from the joint density matrix ${\vec \rho _{[ k -1 ]}}$ of $k-1$ photons and system to the joint density matrix ${\vec \rho _{[ k ]}}$ of $k$ photons and system:
\begin{equation}
{\vec \rho _{[ k ]}} = \sum\limits_{{i_k},{j_k} = 0}^{d-1} {N_{[ k ]}^{{i_k},{j_k}}{{\vec \rho }_{[ {k-1} ]}}} ,\qquad {\rm{  with}}\quad N_{[ k ]}^{{i_k},{j_k}} = W_{\rm ph}^{{i_k}{j_k}}{W_{{{\cal L}_{[ k ]}}}}.
\end{equation}

The fidelity ${{\cal F}_{\rm MPS}}$ can be efficiently evaluated as \cite{zyryd}
\begin{widetext}
\begin{equation} \label{fid_exp}
\begin{array}{l}
{{\cal F}_{{\rm{MPS}}}} = \sum\limits_{\{ {{i_k},{j_k}} \} = 0}^1 {{\rm{Tr}}[ {N_{[ n ]}^{{i_n},{j_n}}...N_{[ 1 ]}^{{i_1},{j_1}}\tilde B} ]} 
 \times {\rm{Tr}}[ {( {A_{[ n ]}^{{j_n}} \otimes \bar A_{[ n ]}^{{i_n}}} )...( {A_{[ 1 ]}^{{j_1}} \otimes \bar A_{[ 1 ]}^{{i_1}}} )( {B \otimes B} )} ],
\end{array}
\end{equation}
\end{widetext}
where we denote $\tilde B \equiv \sum\limits_{\alpha  = 1}^{N_h} {| {{\varphi' _I}} \rangle \langle \alpha | \otimes | {{\varphi' _I}} \rangle \langle \alpha |} $, $B \equiv | {{\varphi _I}} \rangle \langle {{\varphi _F}} |$, and $\bar A$ as the complex conjugate of a matrix $A$. For higher-dimensional rp-PEPS, the ${{\mathcal F}_{{\rm{rp}}}}$ can be computed in the same way by viewing the high-dimensional rp-PEPS as an MPS with bond dimension and physical dimension scale exponentially with the number of sequential photon sources $m$.

\section{Construction of process map $W_{\rm ph}$}
\label{apd_errors}
The decoherence effects and the finite photon retrieval efficiency ${p_{\rm em}}$ during the photon emission will modify $W_{\rm ph}$ from its ideal form ${W_{\rm ph}} = {M_{\rm ph}} \otimes {\bar M_{\rm ph}}$ with $M_{\rm ph}$ of the form \cref{ap_map}. A good way to construct $W_{\rm ph}$ is to include environmental photon modes which capture the erroneous jump events. When there is a finite photon retrieval efficiency ${p_{\rm em}}$, we can include an environmental photon mode ${\varepsilon _T}$, and the ${M_{\rm ph}}:{{\cal H}_T} \to {{\cal H}_T} \otimes {{\cal H}_{\rm ph}} \otimes {{\cal H}_{{\varepsilon _T}}}$ becomes
\begin{equation}
{M_{\rm ph}}:\begin{array}{*{20}{l}}
{{{| 1 \rangle }_T} \to {{| 0 \rangle }_T}( {\sqrt {{p_{{\rm{em}}}}} {{| 1 \rangle }_{{\rm{\rm ph}}}}{{| 0 \rangle }_{{\varepsilon _T}}} + \sqrt {1 - {p_{{\rm{em}}}}} {{| 0 \rangle }_{{\rm{\rm ph}}}}{{| 1 \rangle }_{{\varepsilon _T}}}} )},\\
{{{| 0 \rangle }_T} \to {{| 0 \rangle }_T}{{| 0 \rangle }_{{\rm{\rm ph}}}}{{| 0 \rangle }_{{\varepsilon _T}}}.}
\end{array}
\end{equation}
Here the label $_{\rm ph}$ marks the desired photon mode, and ${\varepsilon _T}$ is an environmental mode that marks the erroneously emitted photon. We construct ${W_{\rm ph}}$ by ${W_{\rm ph}} = {\rm Tr}_{{{\varepsilon _T}}}[ {{M_{\rm ph}} \otimes \bar M_{\rm ph} } ]$, in which we trace out the ${\varepsilon _T}$ mode.

The effect of a transmon decay ${\Gamma _T}$ similarly leads to a branching of the emission, that ${M_{\rm ph}}:{{\cal H}_T} \to {{\cal H}_T} \otimes {{\cal H}_{\rm ph}} \otimes {{\cal H}_{{\varepsilon _T}}}$ becomes 
\begin{widetext}
\begin{equation} \label{eq:1}
{M_{\rm ph}}:\begin{array}{*{20}{l}}
{{{| 1 \rangle }_T} \to {{| 0 \rangle }_T}\left( {\sqrt {\dfrac{{{\Gamma _{{\rm{em}}}}}}{{{\Gamma _{{\rm{em}}}} + {\Gamma _T}}}} {{| 1 \rangle }_{{\rm{\rm ph}}}}{{| 0 \rangle }_{{\varepsilon _T}}} + \sqrt {\dfrac{{{\Gamma _T}}}{{{\Gamma _{{\rm{em}}}} + {\Gamma _T}}}} {{| 0 \rangle }_{{\rm{\rm ph}}}}{{| 1 \rangle }_{{\varepsilon _T}}}} \right)}\\
{{{| 0 \rangle }_T} \to {{| 0 \rangle }_T}{{| 0 \rangle }_{{\rm{\rm ph}}}}{{| 0 \rangle }_{{\varepsilon _T}}}.}
\end{array}
\end{equation}
\end{widetext}

The transmon dephasing leads to an exponential decay of the density matrix elements that are off-diagonal on the transmon basis with rate ${\Gamma _\phi }/2$. In the regime of ${\Gamma _\phi } \ll {\Gamma _{\rm em}}$ the probability accumulates as 
\begin{equation} \label{eq:1}
\int_0^{{T_{{\rm{em}}}}} {\dfrac{{{\Gamma _\phi }}}{2}\langle {{\sigma _{eg}}( t )} \rangle dt}  \approx \int_0^\infty  {\dfrac{{{\Gamma _\phi }}}{2}{e^{- {\Gamma _{{\rm{em}}}}t/2}}dt}  = {\Gamma _\phi }/{\Gamma _{{\rm{em}}}}.
\end{equation}
Thus, the mapping will lead to 
\begin{equation} \label{dep_em}
W_{\rm ph}^{10} \to \left( {1 -\dfrac{{{\Gamma _\phi }}}{{{\Gamma _{\rm em}}}}} \right)W_{\rm ph}^{10}, 	\qquad W_{\rm ph}^{01} \to \left( {1 -\dfrac{{{\Gamma _\phi }}}{{{\Gamma _{\rm em}}}}} \right)W_{\rm ph}^{01}.
\end{equation}
in \cref{wp_map}.

The cavity decay will explicitly depend on the photon emission time ${T_{\rm em}}$. Since ${\Gamma _C} \ll {\Gamma _{\rm em}}$, we can solve the dynamics analytically using the quantum trajectory approach \cite{Plenio1998} and include up to one jump process of the cavity photon. During the emission process,  the decay probability for the Fock state $| n \rangle $ of the cavity mode is approximately $ n{\Gamma _C}{T_{\rm em}}$. By including a cavity decay environmental mode ${\varepsilon _C}$, we can write ${M_{\rm ph}}:{{\cal H}_{C}} \otimes {{\cal H}_{T}} \to {{\cal H}_{C}} \otimes {{\cal H}_{T}} \otimes {{\cal H}_{\rm ph}} \otimes {{\cal H}_{{\varepsilon _C}}}$ as
\begin{widetext}
\begin{equation} \label{eq:1}
{M_{\rm ph}}:\begin{array}{*{20}{c}}
{{{| n \rangle }_C}{{| 1 \rangle }_T} \to \sqrt {1 -n{\Gamma _C}{T_{\rm em}}} {{| n \rangle }_C}{{| 0 \rangle }_T}{{| 1 \rangle }_{{\rm{ph}}}}{{| 0 \rangle }_{{\varepsilon _C}}} + \sqrt {n{\Gamma _C}{T_{\rm em}}} {{| {n-1} \rangle }_C}{{| 0 \rangle }_T}{{| 1 \rangle }_{{\rm{ph}}}}{{| 1 \rangle }_{{\varepsilon _C}}},}\\
{{{| n \rangle }_C}{{| 0 \rangle }_T} \to \sqrt {1 -n{\Gamma _C}{T_{\rm em}}} {{| n \rangle }_C}{{| 0 \rangle }_T}{{| 0 \rangle }_{{\rm{ph}}}}{{| 0 \rangle }_{{\varepsilon _C}}} + \sqrt {n{\Gamma _C}{T_{\rm em}}} {{| {n-1} \rangle }_C}{{| 0 \rangle }_T}{{| 0 \rangle }_{{\rm{ph}}}}{{| 1 \rangle }_{{\varepsilon _C}}}.}
\end{array}
\end{equation}
\end{widetext}
Similarly, we obtain $W_{\rm ph}$ by ${W_{\rm ph}} = {\rm Tr}_{{{\varepsilon _C}}}[ {{M_{\rm ph}} \otimes M_{\rm ph}^\dag } ]$. A finite photon emission time also lead to a residue population on ${p_{1T}}( {{T_{\rm em}}} ) = {e^{ -{\Gamma _{\rm em}}{T_{\rm em}}}}{p_{1T}}( 0 )$ on the state ${| 1 \rangle _T}$. This can be modeled by a further reduction factor on $p_{\rm em}$ that ${p_{{\rm{em}}}} \to ( {1 -{e^{ -{\Gamma _{\rm em}}{T_{\rm em}}}}} ){p_{{\rm{em}}}}$.

With the above analysis, we can write down the ${M_{\rm ph}}:{{\cal H}_{C}} \otimes {{\cal H}_{T}} \to {{\cal H}_{C}} \otimes {{\cal H}_{T}} \otimes {{\cal H}_{\rm ph}} \otimes {{\cal H}_{{\varepsilon _T}}}\otimes {{\cal H}_{{\varepsilon _C}}}$ that includes all the above effects as

\begin{widetext}
\begin{equation}
\begin{array}{l}
	
\begin{aligned}
{| n \rangle _C}{| 1 \rangle _T} \to \sqrt {1 -n{\Gamma _C}{T_{\rm em}}} {| n \rangle _C}{| 0 \rangle _T}\cdot \left( {\begin{array}{*{20}{l}}
{\sqrt {\dfrac{{( {1 -{e^{ -{\Gamma _{\rm em}}{T_{\rm em}}}}} ){\Gamma _{{\rm{em}}}}{p_{{\rm{em}}}}}}{{{\Gamma _{{\rm{em}}}} + {\Gamma _T}}}} {{| 1 \rangle }_{{\rm{ph}}}}{{| 0 \rangle }_{{\varepsilon _T}}}{{| 0 \rangle }_{{\varepsilon _C}}}}
{ + \sqrt {1 -\dfrac{{( {1 -{e^{ -{\Gamma _{\rm em}}{T_{\rm em}}}}} ){\Gamma _{{\rm{em}}}}{p_{{\rm{em}}}}}}{{{\Gamma _{{\rm{em}}}} + {\Gamma _T}}}} {{| 0 \rangle }_{{\rm{ph}}}}{{| 1 \rangle }_{{\varepsilon _T}}}{{| 0 \rangle }_{{\varepsilon _C}}}}
\end{array}} \right)\\
 + \sqrt {n{\Gamma _C}{T_{\rm em}}} {| {n-1} \rangle _C}{| 0 \rangle _T}\cdot \left( {\begin{array}{*{20}{l}}
{\sqrt {\dfrac{{( {1 -{e^{ -{\Gamma _{\rm em}}{T_{\rm em}}}}} ){\Gamma _{{\rm{em}}}}{p_{{\rm{em}}}}}}{{{\Gamma _{{\rm{em}}}} + {\Gamma _T}}}} {{| 1 \rangle }_{{\rm{ph}}}}{{| 0 \rangle }_{{\varepsilon _T}}}{{| 1 \rangle }_{{\varepsilon _C}}}}
{ + \sqrt {1 -\dfrac{{( {1 -{e^{ -{\Gamma _{\rm em}}{T_{\rm em}}}}} ){\Gamma _{{\rm{em}}}}{p_{{\rm{em}}}}}}{{{\Gamma _{{\rm{em}}}} + {\Gamma _T}}}} {{| 0 \rangle }_{{\rm{ph}}}}{{| 1 \rangle }_{{\varepsilon _T}}}{{| 1 \rangle }_{{\varepsilon _C}}}}
\end{array}} \right),
\end{aligned}
\\
\\
{| n \rangle _C}{| 0 \rangle _T} \to \sqrt {1 -n{\Gamma _C}{T_{\rm em}}} {| n \rangle _C}{| 0 \rangle _T}{| 0 \rangle _{{\rm{ph}}}}{| 0 \rangle _{{\varepsilon _T}}}{| 0 \rangle _{{\varepsilon _C}}} + \sqrt {n{\Gamma _C}{T_{\rm em}}} {| {n-1} \rangle _C}{| 0 \rangle _T}{| 0 \rangle _{{\rm{ph}}}}{| 0 \rangle _{{\varepsilon _T}}}{| 1 \rangle _{{\varepsilon _C}}}.
\end{array}
\end{equation}

\end{widetext}

\begin{figure*}
	\centering
	\includegraphics[width=0.95\textwidth]{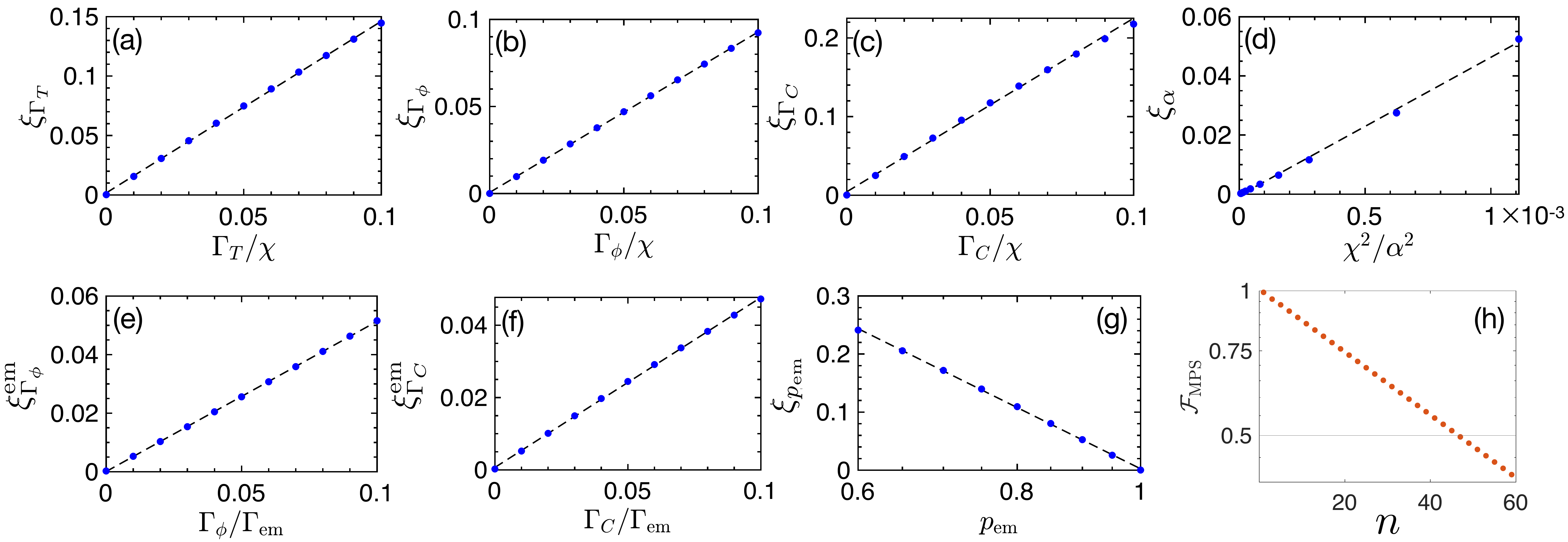}
        \caption{Scaling of the error rate $\xi$ in ${{\mathcal F}_{{\rm{\rm MPS}}}}$ for creating the cluster state [\cref{cls_form}] using pulse sequence in \cref{qoc_cq} as a function of: the transmon decay during unitary operation (a), transmon dephasing during unitary operation (b), cavity decay during unitary operation (c), transmon anharmonicity during unitary operation (d), transmon dephasing during photon emission (e), cavity decay during photon emission (f), and the photon retrieval efficiency (g). The panel (h) show the MPDO calculation of ${{\mathcal F}_{{\rm{\rm MPS}}}}$ versus the cluster state photon number $n$, with transmon parameters in \cref{fid_eth}. The horizontal line denotes ${\cal F} _{\rm MPS}=1/2$.}
        \label{add_scaling}
\end{figure*}

And further trace out the environmental modes to get ${W_{\rm ph}} = {\rm Tr}_{{{\varepsilon _T},{\varepsilon _C}}}[ {{M_{\rm ph}} \otimes \bar M_{\rm ph} } ]$. After that, we include the transmon dephasing by applying \cref{dep_em}, to finish the construction of ${W_{\rm ph}}$.

\section{Additional scaling data for MPS preparation fidelity}
\label{beta_scale}

\subsection{Scaling of the error rate $\chi$}
As shown in the main text, we can compute the scaling of the coefficient $\xi$ of the exponentially decaying fidelity ${\cal F}_{\rm MPS} = e^{-\xi \cdot n}$ with the slope (error rate) $\xi$ as a function of various imperfections. This scaling is numerically shown here in \cref{add_scaling}(a-g), where each data point is extracted from a MPDO calculation of the relation between ${\cal F}_{\rm MPS}$ and the photon number $n$ (an example is shown in \cref{pe_plot}a), with only the specific decoherence channel turned on. For example, in \cref{add_scaling}(a) only the transmon decay ${\Gamma _T}$ is turned on. From \cref{add_scaling}(a-g) we obtain all individual terms in \cref{fid_scale}, with non-universal coefficients in \cref{scale_beta} for the cluster state generation with the pulse sequence in \cref{qoc_cq}. In the regime where the error per photon generation is small, we can estimate the total error $\xi$ by simply adding these individual terms, thus obtaining \cref{fid_scale}.

\subsection{Fidelity versus photon number for transmon parameters in Ref.~\cite{Besse2020}}
\label{fid_eth}
In \cref{pe_plot}(a) we showed the fidelity versus the cluster state photon number for the state-of-the-art experimental parameters shown in \cref{mps_perf}. To better compare it with the experimentally demonstrated result~\cite{Besse2020}, here we compute the same relation with the transmon properties reported there, while other parameters stays the same as that in \cref{mps_perf}. Specifically, we choose ${\Gamma _T} = \SI{47.62}{\rm kHz}$, ${\Gamma _\phi } = \SI{58.82}{\rm kHz}$, 
$\alpha  = 2\pi  \times - \SI{303}{\rm MHz}$, and $\chi  = {\rm{2}}\pi  \times  - \SI{5}{\rm MHz}$~\cite{Besse2020}. The result is shown in \cref{add_scaling}(h), from which we get $N_{\rm ph}\approx 47$.

\subsection{Achievable entanglement length $N_{\rm ph}$ with MPS bond dimension $D$}
\label{mps_bond}

The entanglement length ${N_{{\rm{ph}}}} = \log 2/\xi $ is determined by the error rate per photon $\xi$ [\cref{fid_scale}]. To produce an MPS with bond dimension $D$, we need to implement unitaries on $2D$-dimensional Hilbert space. As numerical evidence suggesting that \cite{Lee2018b} the time cost of implementing a general unitary in $N$-dimensional Hilbert space using the quantum optimal control approach scales as $O(N^2)$, it takes $T^D_{\rm MPS}=O(D^2)$ to implement above unitaries. This lead to increased decoherence as the coefficients ${\beta _C},{\beta _T },{\beta _\phi},{\beta _\alpha}$ in \cref{err_scale} are proportional to $T^D_{\rm MPS}$. In the regime of ${p_{{\rm{em}}}} \approx 1$ and $\beta _0 \approx 0$ (typical for current experimental platforms~\cite{Heeres2017,Besse2020}), 
one can thus estimate the scaling of $\xi$ as
\begin{equation} \label{err_opt_bond}
{\xi } \approx T_{{\rm{MPS}}}^D{\xi _{{\rm{unit}}}} + \xi _{{\rm{em}}}^{{\rm{src}}} + \xi _{{\rm{em}}}^{{\rm{ph}}}\sim O( {{D^2}} ).
\end{equation} 

Thus the dominant part lead to a qualitative scaling of ${N_{{\rm{ph}}}}\sim {D^{ - 2}}$.

\section{Isometric tensor network states}
\label{iso_psg}

In this section, we provide more details on the definition of isometric tensor network states (isoTNS)~\cite{Zaletel2020}.

\begin{figure}[h!]
	\centering
	\includegraphics[width=0.48\textwidth]{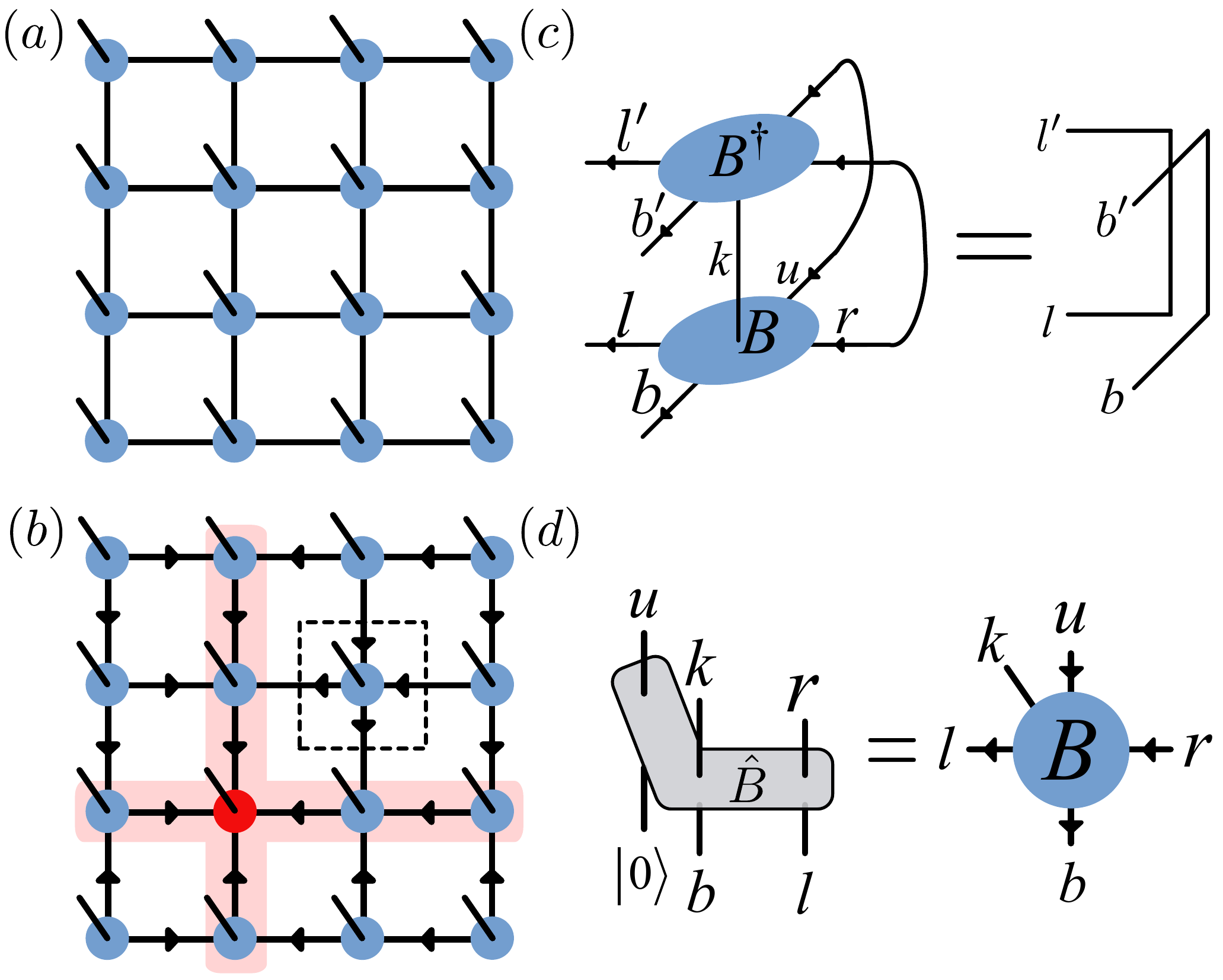}
        \caption{The projected entangled pair states (PEPS) and isometric tensor network states (isoTNS). (a) PEPS are states represented by a network of tensors, where the tensor at each site is of physical dimension $d$, and is connected to that of the neighboring sites with virtual bonds of dimension $D$. (b) IsoTNS are a subclass of PEPS, where the tensors satisfy isometry conditions denoted by the arrows. The red (gray) shaded lines denote the \textit{orthogonality hypersurface} of isoTNS, and their crossing point [the red (dark gray) dot] denote the \textit{orthogonality center} of isoTNS. (c) The isometry condition for the tensor inside the dashed box in panel (b). Here the incoming virtual legs and the physical legs of the tensor are contracted with the corresponding legs of the complex conjugate of the tensor, yield identity on the outgoing legs. This is also mathematically shown in \cref{isop}. (d) One can map a three-qubit `L'-shaped quantum gate (see \cref{isop_tori}) to an isoTNS tensor of bond dimension $D=d$.}
        \label{peps_iso}
\end{figure}

To start, first we recall the definition of the projected entangled pair states (PEPS)~\cite{verstraete2004renormalization}, which are defined through a network of tensors that are connected with each other, with one tensor at each lattice site (see \cref{peps_iso}a). The wavefunction of PEPS is obtained by contracting the connected (virtual) legs of the tensors, as
\begin{equation} \label{}
	\left|\Psi_{\mathrm{PEPS}}\right\rangle=\sum_{\{k\}=0}^{d-1} \mathcal{F}_{2 \mathrm{D}}(\{B_{[i, j]_{l u r b}}^{k}\})|\{k\}\rangle,
\end{equation}
where the ${B_{\left[ {i,j} \right]lurd}^k}$ is a rank-5 tensor on the site $(i,j)$, which has virtual indices $l.u,r,b$ of bond dimension $D$ and physical index $k$ of physical dimension $d$. And the symbol ${\cal F}_{2D}$ denote the contraction of the connected virtual indices. PEPS serve as a natural extension of MPS in higher dimensions, and has wide applications in describing higher-dimensional many body systems~\cite{orus2014practical,Orus2019,Cirac2020}.

IsoTNS is a subclass of PEPS, where the tensors satisfy certain isometry conditions. The isometry condition means that, when the incoming legs (denoted by the arrows in \cref{peps_iso}b) and the physical legs of a tensor are contracted with corresponding legs of the complex conjugate of this tensor, the remaining legs yield an identity. For example, the tensor in the dashed box in \cref{peps_iso}(b) obeys
\begin{equation} \label{isop}
\sum\limits_{k,ur} {{B_{[ {i,j} ]}^k}_{lurb}{{( {{B_{[ {i,j} ]}^k}_{l'urb'}} )}^*}}  = {\delta _{bb'}}{\delta _{ll'}},
\end{equation}
which is shown graphically in \cref{peps_iso}(c). Moreover, the two red shaded lines in \cref{peps_iso}(b) only have incoming arrows, which are termed the \textit{orthogonality hypersurface} of isoTNS, and their intersection is the \textit{orthogonality center} of the isoTNS~\cite{Zaletel2020}.

As shown in Ref.~\cite{zypp}, one can convert an isoTNS tensor $B_{\left[ {i,j} \right]lurb}^k$ of the bond dimension $D$ into a `L'-shaped unitary of the form
\begin{equation} \label{iso_unit_atom}
    {\hat B_{[ {i,j} ]}} = \sum\limits_{lurb,k} {{B_{[ {i,j} ]}^k}_{lurb}} {| {u,k,r} \rangle }\langle {0,b,l} |,
\end{equation}
where the indices are identified in \cref{peps_iso}(d) for the case of bond dimension $D=d$. Here the rank-5 tensor ${{B_{[ {i,j} ]}^k}_{lurb}}$ also satisfy the same isometry condition \cref{isop}. This show that the gates in the photon generation scheme shown in \cref{isop_tori} can be identified as isoTNS tensors. Increasing isoTNS bond dimension corresponds to increasing the arm length of the `L'-shaped unitary.

In this way, one can generate isoTNS by sequentially applying overlapping `L'-shape unitaries, which lead to the photonic isoTNS generation protocol discussed in \cref{iso_ex}. This fact further implies that isoTNS is a subclass of rp-PEPS~\cite{zypp} since we can cover the `L'-shape unitaries by plaquette unitaries.

\subsection{IsoTNS representation of the toric code}
\label{tc_iso}
In general, the ${\mathbb Z}_\lambda$ toric code can be written as an isoTNS with the translational-invariant tensor at each site as \cite{Schuch2010}
\begin{equation} \label{toric_d}
B_{lurd}^{{i_1}{i_2}{i_3}{i_4}} = \frac{1}{\lambda}\delta _{l - u}^{{i_1}}\delta _{u - r}^{{i_2}}\delta _{r - b}^{{i_3}}\delta _{b - l}^{{i_4}},
\end{equation}
where
\begin{equation} \label{}
\delta _b^a = \left\{ \begin{array}{l}
1,\qquad a = b\,\,\bmod \,\,\lambda\\
0,\qquad {\rm{otherwise.}}
\end{array} \right.
\end{equation}
This is a tensor of bond dimension $\lambda$ and physical dimension $\lambda^4$, and satisfies the isometry condition \cref{isop}.

\section{Control universality of $H_{\rm{array}}(t)$}
\label{apd_univ}
Here we show the Hamiltonian $H_{\rm{array}}(t)$ [\cref{peps_ham_tot}] can universally control the Hilbert space ${{\cal H}^{L_C \times m}_{\rm array}} = {( {{{\cal H}_{\rm src}}} )^{ \otimes L_C \times m}}$. Let us consider case of two cQED sequential photon sources coupled to each other ($L_C=1$ and $m=2$), that the whole Hilbert space is 
\begin{equation} \label{}
	{{\cal H}^{1\times 2}_{\rm array}} = {{\cal H}_T} \otimes {{\cal H}_C} \otimes {{\cal H}_C} \otimes {{\cal H}_T}.
\end{equation}

We know the Hamiltonian for each sequential photon source can control one ${{{\cal H}_{\rm src}}}$ universally. This means one can create arbitrary Hamiltonians that act on each Hilbert space ${{\cal H}_{C}}$ of the cavity mode (see \cref{poly_ham_cav} below). Together with the bilinear interaction $H_{\rm int}^{ij}$ [\cref{bilin_ham}] between two cavities, one can apply an analogous argument in Ref.~\cite{Lloyd1999}, that by arithmetic operation and taking commutators between the bilinear coupling $H_{\rm int}^{12}$ [cf.~\cref{bilin_ham}] and single cavity Hamiltonians
\begin{equation} \label{poly_ham_cav}
{H_1} = {\rm poly}( {{a_1},a_1^\dag } ),\quad{H_2} = {\rm poly}( {{a_2},a_2^\dag } ),
\end{equation}
we can generate arbitrary polynomial form of Hamiltonian 
\begin{equation} \label{}
{H_{12}} = {\rm poly}( {{a_1},a_1^\dag ,{a_2},a_2^\dag } ) \in {{\cal H}_C} \otimes {{\cal H}_C}.
\end{equation}
This immediately implies that we can universally control the Hilbert space of two cavities \cite{Lloyd1999}. Together with universal control on each sequential photon source, we can use Lemma 5.5 of Ref. \cite{Hofmann2017} to combine the universality of two ${{\cal H}_{\rm src}}$ and ${{\cal H}_C} \otimes {{\cal H}_C}$ to the whole ${{\cal H}^{1\times 2}_{\rm array}}$, which shows that we can universally control ${{\cal H}^{1\times 2}_{\rm array}}$ with $H_{\rm{array}}(t)$ (here $L_C=1$ and $m=2$). By repeatedly applying Lemma 5.5 of Ref. \cite{Hofmann2017}, one can show $H_{\rm{array}}(t)$ is further able to control ${{\cal H}^{L_C\times m}_{\rm array}}$.

\bibliography{/Users/gishiyuan/Documents/library.bib}

\end{document}